\documentclass[a4paper,11pt]{article}
\pdfoutput=1 

\usepackage{jheppub} 

\usepackage[T1]{fontenc} 
\usepackage{hyperref}
\usepackage{braket}
\usepackage{stackengine}
\usepackage{xcolor}
\usepackage{graphicx}
\usepackage{subfigure}
\title{\boldmath Geometric quantum complexity of bosonic oscillator systems}

\author[a,1]{Satyaki Chowdhury,\note{Corresponding author.}}
\author[b]{Martin Bojowald,}
\author[a]{Jakub Mielczarek}

\affiliation[a]{Institute of Theoretical Physics, Jagiellonian University, Lojasiewicza 11, 30-348 Cracow, Poland}
\affiliation[b]{Institute for Gravitation and the Cosmos, The Pennsylvania State University,
104 Davey Lab, University Park, PA 16802, USA}

\emailAdd{satyaki.chowdhury@doctoral.uj.edu.pl}
\emailAdd{bojowald@psu.edu}
\emailAdd{jakub.mielczarek@uj.edu.pl}

\abstract{According to the pioneering work of Nielsen and collaborators, 
the length of the minimal geodesic in a geometric realization of a suitable operator space provides a 
measure of the quantum complexity of an operation. Compared with the original concept of complexity based on the minimal number of gates required to construct the desired operation as a product, this geometrical approach amounts to a more concrete and computable definition, but its evaluation is nontrivial in systems with a high-dimensional Hilbert space. The geometrical formulation can more easily be evaluated by considering the geometry associated with a suitable finite-dimensional group generated by a small number of relevant operators of the system. In this way, the method has been applied in particular to the harmonic oscillator, which is also of interest in the present paper. However, subtle and previously unrecognized issues of group theory can lead to unforeseen complications, motivating a new formulation that remains on the level of the underlying Lie algebras for most of the required steps. Novel insights about complexity can thereby be found in a low-dimensional setting, with the potential of systematic extensions to higher dimensions as well as interactions. Specific examples include the quantum complexity of various target unitary operators associated with a harmonic oscillator, inverted harmonic oscillator, and coupled harmonic oscillators. The generality of this approach is demonstrated by an application to an anharmonic oscillator with a cubic term. 
}
\begin{document} 
\maketitle
\flushbottom

\section{Introduction}

Quantum complexity, a notion of measuring the difficulty of carrying 
out a complicated task as a succession of simple operations, is an 
important concept in a variety of different areas. For instance, it 
has surprisingly helped to study important physical problems such as 
information processing by black holes. In this context, the famous 
Ryu-Takayanagi proposal \cite{Nishioka:2009un,Lewkowycz:2013nqa,Rangamani:2016dms} 
of relating entanglement entropy (between two regions in the boundary 
conformal field theory) to minimal surfaces in AdS spaces proved 
to be extremely successful in interpreting spacetimes as emergent 
from quantum entanglement. Black holes posed serious challenges 
because the volume behind the horizon of the AdS black hole continues 
to grow for a very long time well after the saturation of its entanglement 
entropy \cite{Hartman:2013qma}. Hence, it became essential to look 
for quantities that capture this long-term growth. Susskind and 
collaborators'  \cite{Susskind:2014jwa,Susskind:2014moa,Susskind:2014rva} 
suggested that the growth of the volume should be dual to the quantum 
complexity of the boundary field theory that led to the ``Complexity=Volume'' 
(CV) conjecture \cite{Susskind:2014jwa,Susskind:2014moa,Susskind:2014rva}, 
followed by the ``Complexity=Action'' (CA) conjecture, which related complexity 
of the field theory to the gravitational action of the Wheeler-de-Witt patch 
of the bulk gravity theory \cite{Brown:2015bva,Brown:2015lvg}. The most recent 
viewpoint of complexity is known as the ``Complexity=Anything'' \cite{Belin:2021bga,Belin:2022xmt}.

In quantum gravity, the motivation to study quantum complexity is primarily derived from the enigmatic 
black hole phenomena. Nevertheless, its importance extends to numerous other 
branches of physics, particularly those involving complex systems\footnote{A 
comprehensive review of recent advancements in quantum complexity across the 
field of physics is presented in \cite{Chapman:2021jbh}.}. In broad terms, a complex system is generally characterized by a high number of degrees of freedom, 
non-linear interactions, and long-distance correlations. There exists a plethora 
of complex systems across various disciplines of physics, ranging from strongly 
correlated quantum systems to biological systems and even extending to social networks. 
In this discussion, however, our focus is specifically limited to systems that 
demonstrate quantum properties.

The notion of quantum complexity is relatively new in physics, but in computer 
science and information theory various relevant notions of complexity have been 
used for some time. All of them have something to do with how a complex object 
is built from simple or small constituents, and how many of these simple parts 
are needed to make the complex whole. Two of the most popular notions of complexity 
in computer science are the Kolmogorov complexity \cite{Ming} and the stochastic 
complexity \cite{Rissanen}. The Kolmogorov complexity of an object measures 
the length of the shortest program required to produce it.  The stochastic complexity 
is applied to messages and the smallest code required to compress them. The 
general notion of complexity was inherited from computer science and information 
theory by quantum computation research in efforts to determine the complexity 
of quantum circuits.

In a groundbreaking series of papers, Nielsen and his collaborators \cite{Nielsen_2006,https://doi.org/10.48550/arxiv.quant-ph/0502070,https://doi.org/10.48550/arxiv.quant-ph/0701004} established a correlation 
between quantum complexity and the length of the minimal geodesic in the space of 
unitary operators. This geometric definition opened up an entirely new avenue 
of research, leading to significant advancements in understanding the role of 
quantum complexity in holography. Nielsen's formulation primarily focused on determining the minimal circuit size necessary to implement an $n$-qubit unitary operation, thereby determining the 
complexity of specific quantum states. However, its significance in the broader context 
of quantum mechanics remains largely unexplored. Several notable efforts have recently 
been made to 
quantify the complexity of individual states \cite{Jefferson:2017sdb,Chapman:2018hou,Caceres:2019pgf, Khan:2018rzm, Hackl:2018ptj,Chapman:2017rqy,Bhattacharyya:2018bbv, Bhargava:2020fhl,Adhikari:2021ckk, Guo:2018kzl}.

A path integral optimization technique to define complexity in CFTs were developed in \cite{Caputa:2017urj,Caputa:2017yrh}. It was later shown in \cite{Camargo:2019isp} that these results can be interpreted as a version of Nielsen's circuit complexity. Similarly, in \cite{Jefferson:2017sdb,Chapman:2017rqy}, attempts were made to define complexity in quantum field theories. For example, \cite{Jefferson:2017sdb}
was interested in the complexity of the ground state in free scalar field theory. For this purpose, choosing a suitable reference state is required. The approach of state complexity is then dependent on the choice of reference state. In practice, many of these attempts 
are limited by the requirement that the reference and target states be Gaussian.  Several methods were developed and extensively applied to deal with the complexity of Gaussian states, like the covariance matrix method \cite{Chapman:2018hou}, 
Fubini--Study method \cite{Chapman:2017rqy}. An attempt to analyze complexity for 
a non-Gaussian state was made in \cite{Bhattacharyya:2018bbv}. A comparative 
study of different approaches to state complexity was carried out in \cite{Ali:2018fcz}. While these achievements represent remarkable 
progress, they do not provide a comprehensive understanding of the 
complexity associated with quantum processes in general. 

In our studies, we will be primarily interested in operator complexity, 
specifically in the complexity of time evolution operators associated 
with certain Hamiltonians rather than the complexity of a certain state. 
This article has two main motivations:
\begin{itemize}
    \item Providing a general recipe to determine the complexity of any unitary operator without needing a separate definition of reference or target states. In the literature, the notion of operator complexity is mainly introduced by characterizing their action on Gaussian reference and target states. These limitations make it challenging to provide general statements on the complexity of the operator. A general state-independent approach is expected to be more feasible in extensions of the analysis to interacting systems or in applications to Hamiltonians beyond the quadratic level.
    \item Understanding how the complexity changes as a quantum system undergoes time evolution. For this purpose, the target unitary operators, whose complexity we will be interested in, are time-evolution operators for various systems.
\end{itemize}
The determination of the complexity of a certain state has its significance in the context of quantum simulations, particularly in many body systems and their wave functions. An important example is the complexity of specific states, such as the thermofield double state, which is conjectured to be dual to an eternal black hole in anti-de Sitter space-time and may, therefore, play a crucial role in the understanding of stationary black holes. By contrast, the complexity of time evolution addresses the question of the least complex way of evolving a quantum system: Is it possible to find a shorter path between two points in the evolution other than the one generated by the Hamiltonian itself?
For our exploration, we will consider some of the widely considered models, including those considered in \cite{Jefferson:2017sdb,Chapman:2018hou,Ali:2019zcj}. However, we will not be interested in the complexity of a particular state but rather in the complexity of the time-evolution operator. 

Nielsen's general method, which can be applied to non-qubit systems by replacing 
the unitary groups ${\rm SU}(N)$ with other suitable Lie groups, is well-suited 
to this purpose. If the quantum system describes the motion of a point particle, 
the method requires a truncation of the space of all unitary operators on an 
infinite-dimensional Hilbert space to a suitable finite-dimensional group. In the 
case of the harmonic oscillator, with its wide-ranging applications in various fields, 
the groups ${\rm SU}(1,1)$ and ${\rm Sp}(2,{\mathbb R})$ play a prominent role 
(see for instance \cite{Haque:2021hyw} for such applications to complexity). 
However, unlike ${\rm SU}(N)$, these groups are not compact, leading to mathematical 
subtleties that, to the best of our knowledge, have not been appreciated in 
the existing literature. We will discuss these issues in the present paper 
--- such as the non-existence of finite-dimensional unitary matrix representations, 
non-surjectivity of the exponential map, and geodesic incompleteness of group metrics 
--- and analyze to what degree they can be avoided by working on an algebraic 
rather than group-theoretic level whenever possible. 
We also demonstrate extensions of this method that allow us to go beyond 
quadratic Hamiltonians, providing the first results of complexity for anharmonic systems.

The rest of the paper is organized as follows: in Section \ref{sec:Nielsengeometricmethod}
we review the essentials of quantum complexity. We start by introducing the 
conventional definition of quantum complexity described in terms of the number 
of universal gates in the desired quantum circuit. We explain the significance 
of the geometrical definition of quantum complexity by pointing out the drawbacks 
and limitations of the gate definition of quantum complexity. We then provide 
a quick review of the geometrical approach to quantum complexity with a complete 
recipe for a general target unitary operator. In Section \ref{sec:HOgroup}, we 
briefly discuss the essential quantities corresponding to the harmonic oscillator 
group as a minimal implementation of the geometric method, highlighting several 
subtleties, and illustrate the recipe 
of the geometrical framework of quantum complexity bounds in this system by an explicit 
computation for the displacement operator 
and several examples of harmonic time evolution operators, 
which was the primary motivation of the paper.
In Section \ref{sec:complexitySU}, we show that not only the harmonic oscillator 
group but also the symplectic group ${\rm Sp}(2,{\mathbb R})$ is efficient in determining the complexity 
of the harmonic oscillator, and in addition of the inverted harmonic oscillator 
which cannot be described using the harmonic oscillator group. In Section 
\ref{sec:CoupledHO}, we consider the case of two coupled harmonic oscillators 
and study the complexity of time evolution in such coupled systems. Our final example
discusses an anharmonic oscillator with a cubic term, after which we conclude with 
a brief discussion of our findings and future directions. Various calculations 
details are relegated to appendices. 

\section{Overview of quantum complexity}
\label{sec:Nielsengeometricmethod}

In this section, we briefly review Nielsen's geometric interpretation of 
quantum complexity. As discussed earlier, the notion of quantum complexity 
measures the minimum number of simple operations required to carry out 
a complex task. In terms of unitary operators (or quantum states), 
we might think of complexity as the minimum number of elementary gates 
required to construct the quantum circuit that will produce the desired 
unitary operator (or takes a reference state to the target state). 
This approach of counting gates is better known as \emph{gate complexity}.

\subsection{Gate approach to quantum complexity -- Gate Complexity}

Consider $U_{\rm target}$, which represents the target unitary operator 
whose complexity we are interested in. Actual construction of a quantum 
circuit would proceed by building it out of elementary operations $g_i$ 
that synthesize $U_{\rm target}$:
\begin{equation}
    U_{\rm target} =g_ng_{n-1}\cdots g_2g_1.
\end{equation}
Out of all possible circuits built from the chosen gate set $\{g_i\}$, 
realizing $U_{\rm target}$ (up to a certain level of accuracy $\epsilon$), 
the circuit requiring the minimal number of gates in $\{g_i\}$ is the 
optimal one, and this number of gates in the optimal circuit gives 
a measure of the complexity of $U_{\rm target}$. Therefore,
\begin{align*}
    \textbf{Gate Complexity} \equiv\; &\textbf{number of quantum gates used in the optimal circuit} \\ &\textbf{that implements the desired $U_{\rm target}$ within $\epsilon$.}
\end{align*}

The gate approach of quantifying complexity is adapted to the question 
of how to build an actual quantum circuit out of component gates, but 
this definition is not suitable for quantum systems with continuous variables. 
One of the major drawbacks is the choice of a set of gates. For example, 
if two gate sets \{$g_i$\} and \{$m_i$\} can realize the same 
$U_{\rm target}$, the numbers of gates in the optimal circuit 
required from \{$g_i$\} ($n_1$) and from $\{m_i\}$ ($n_2$) are in 
general different. Therefore, a well-defined complexity of 
$U_{\rm target}$ ($n_1$ or $n_2$) requires an \emph{a priori} 
choice of the gate set, $\{g_i\}$ or $\{m_i\}$. The gate complexity 
is thus not uniquely defined for a target unitary. 

Another significant drawback of gate complexity is its sensitivity 
to the level of accuracy ($\epsilon= |U_{\rm target}-g_ng_{n-1}...g_2g_1|$ 
in some operator norm) required for the quantum circuit. Using discrete 
gates to build a circuit may result in producing a unitary $U^{A}_{\rm target}$, 
which is close to but not exactly equal to the desired $U_{\rm target}$. 
The need to use an operator norm renders this concept of closeness non-unique. In the above 
example, instead of $U_{\rm target}$, we would apply $U^A_{\rm target}$ to 
an initial state of a quantum system. The dynamics produced by $U^A_{\rm target}$ 
is not identical to the desired dynamics, generated by $U_{\rm target}$, and 
will have some correction terms which might play a significant role for 
instance in quantum computation. The correction terms can be reduced by making 
$U^A_{\rm target}$ as close as possible to $U_{\rm target}$, using smaller and 
smaller $\epsilon$, but the outcomes of the actual target operator may be different in each case. Therefore, 
a dependence of complexity on the sensitivity $\epsilon$ is not desirable.  
This level of non-uniqueness in the definition of quantum complexity calls 
for a better and more concrete alternative. A geometrical viewpoint of 
quantum complexity was given by Nielsen and his collaborators, which we 
discuss in the next subsection.

\subsection{Geometrical approach to quantum complexity}

In a series of papers, Nielsen {\em et.\ al\/} \cite{Nielsen_2006,https://doi.org/10.48550/arxiv.quant-ph/0502070,https://doi.org/10.48550/arxiv.quant-ph/0701004} proposed a 
transition from the discrete picture of gate complexity to a continuous 
description, making new connections between quantum complexity and 
differential geometry. They observed that determining the quantum complexity 
of a unitary operation is closely related to the problem of finding minimal-length
geodesics in a certain curved geometry. The original motivation for introducing 
a geometrical notion of quantum complexity was to use it as a tool to bound the 
value of gate complexity. From this initial definition, it has evolved into a candidate for a 
fundamental and unique definition of quantum complexity, a viewpoint that is supported by its smaller degree of ambiguity. Therefore, from now 
on, we will cease thinking of complexity geometry as a continuous approximation 
to gate complexity but rather view gate complexity as a discrete approximation 
of geometric complexity. 

Let us briefly review the overall idea of the geometrical method to quantum 
complexity. In this approach, the complexity of a unitary operator $U$ is the 
length of the minimal geodesic on the unitary group manifold joining the 
identity to $U$. In the original manifestation of the geometric approach, 
unitaries acting on $n$-qubit systems were investigated, and the framework 
relied on the special unitary groups ${\rm SU}(2^n)$. An extension of the 
basic idea of the entire framework to a general unitary is initially 
straightforward, but it does lead to several mathematical subtleties, 
some of which will be described in more detail in the main part of our paper. 
For instance, a general discussion of complexity in quantum mechanics would 
require differential geometry on infinite-dimensional manifolds. Properties 
of geodesics are then hard to analyze, not only because solving an infinite 
set of coupled differential equations is usually difficult, but also because 
a geodesic between two given points is then not guaranteed to exist 
(since the Hopf--Rinow theorem no longer applies). Instead, geodesic distance 
is defined as the infimum (not necessarily a minimum) of the distance on the 
space of all possible curves connecting the two points, and there may be no 
curve that has the resulting distance. For a tractable application of the method, one should therefore, first reduce the infinite-dimensional problem 
to a finite-dimensional one, depending on the target unitary of interest. 
Even then, properties of Lie groups may lead to further subtleties.

Given the target unitary operator whose complexity is of interest, one may 
identify a set of fundamental operators related in some way to it. In particular, 
the Lie algebra generated by a suitable choice of fundamental operators of 
the quantum system (of which there are finitely many ones in quantum mechanics, 
as opposed to quantum field theory) can be exponentiated to a group of which 
the target unitary is an element. If the target unitary is defined as the 
exponential of a Hamiltonian (times $i$), the task in this step is to find 
a suitable set of other operators, including additional observables of interest 
such as position and momentum, that, together with the Hamiltonian, have 
a closed set of commutators. After identifying these generators of a Lie algebra, 
one classifies them as ``easy'' or ``hard'' for an application in the given 
quantum system. For instance, the closure condition on the brackets may require 
one to use additional generators that are physically less motivated than the 
original observables, which accordingly would be considered ``hard'' to construct. 
Their contributions are then suppressed in the geodesic distance by assigning large metric components to their directions. The Lie 
algebra generated by all the operators, easy and hard ones, is exponentiated 
to a Lie group. If the generators are self-adjoint, the Lie group is unitary.

In order to define geometry, one then considers a right-invariant metric 
that accurately captures the hardness property by penalizing the directions 
along the hard operators such that moving in their direction is discouraged 
for geodesics on the Lie group. In the literature on quantum complexity, the set of hardness 
coefficients is known as the \emph{penalty factor} matrix, denoted by $G_{IJ}$. 
The choice of the matrix $G_{IJ}$ is usually motivated by phenomenological 
considerations \cite{Auzzi:2020idm,Bhattacharyya:2019kvj}, inspired by difficulties of performing certain operations 
during an experiment \cite{Brown:2019whu}. Sometimes, theoretical bias is also 
used in this choice. Assuming different penalties or operational costs in 
different directions ($G_{IJ}\neq \delta_{IJ}$) introduces anisotropy in 
the resulting operator space geometry.  The choice of the metric leads to 
a notion of distance on the unitary space, which, as recalled in 
Appendix~\ref{appA}, is given by:
\begin{align}
\label{lineelement}
    ds^2 = \frac{1}{{\rm Tr}(\mathcal{O}_I\mathcal{O}^{\dagger}_I) {\rm Tr}(\mathcal{O}_J\mathcal{O}^{\dagger}_J)}\bigg[G_{IJ} {\rm Tr}[i U^{-1}\mathcal{O}_I^{\dagger}dU] {\rm Tr}[i U^{-1}\mathcal{O}_J^{\dagger}dU]\bigg],
\end{align}
where the $\mathcal{O}_I$ represent the generators of the unitary group and $U$ plays the role of a point on the manifold. The trace ${\rm Tr}$ is taken in a matrix representation of the generators. For geodesics, only the right-invariance of the line element matters, but not the specific form on the entire group.

An efficient way of determining geodesics on Lie groups equipped with a 
right-invariant metric was given by Arnold and is known as the Euler--Arnold 
equation \cite{Balasubramanian:2019wgd}:
\begin{equation}
\label{eqn:eulerarnold}
	G_{IJ}\frac{dV^{J}(s)}{ds}= f_{IJ}^{K} V^{J}(s) G_{KL}V^{L}(s),
\end{equation}
where $f_{IJ}^{K}$ are the structure constants of the Lie algebra, defined by:
\begin{equation}
\label{structure}
	[\mathcal{O}_{I},\mathcal{O}_{J}]= i f_{IJ}^{K} \mathcal{O}_{K}.
\end{equation}

The Euler-Arnold equations have been extensively used recently to compute the 
geodesics on unitary manifolds, see Refs.  \cite{Balasubramanian:2018hsu,Balasubramanian:2019wgd,Balasubramanian:2021mxo, Flory:2020dja, Basteiro:2021ene, Auzzi:2020idm}.

The components $V^I(s)$ represent the tangent vector (or the velocity) at 
each point along the geodesic, defined by:
\begin{align}
\label{differentialU}
	\frac{dU(s)}{ds}=-i V^I(s)\mathcal{O}_I U(s).
\end{align}
The coupled differential equations (\ref{eqn:eulerarnold}) do not depend 
on the position $U(s)$ along the geodesic, and can therefore be solved 
independently of (\ref{differentialU}). Given a solution $V^I(s)$, a further 
integration of (\ref{differentialU}) results in the path (or trajectory) 
in the group, guided by the velocity vector $V^I(s)$. Generically, this 
solution can be written as  the path-ordered exponential:
\begin{equation}
\label{sdependentunitary}
	 U(s)=\mathcal{P}\exp\bigg(-i\int_{0}^{s}ds'~V^I(s')\mathcal{O}_I\bigg),
\end{equation}
on which we impose the boundary conditions 
\begin{equation}
\label{boundary}
	U(s=0)=\mathbb{I} ~~~~{\rm and}~~~ U(s=1)=U_{\rm target},
\end{equation}
where $U_{\rm target}$ is some target unitary whose complexity we wish to study. 
Implementing the boundary condition $U(1)=U_{\rm target}$ in 
order to derive the $v_i$ for a specified target unitary operator requires 
dealing with the path-ordered exponential (\ref{sdependentunitary}), which 
is a notoriously tricky task. The usual way of deriving it is an iterative 
approach expressed as a Dyson series:
\begin{align}
    U(s)= \mathbb{I}-i \int_{0}^{s}V^I(s')O_I ds' + (-i)^2 \int_{0}^{s} V^I(s') O_I ds'\int_{0}^{s'}V^I(s'')O_I ds''+\cdots. 
    \label{DysonSeries}
\end{align}

In general, equation (\ref{eqn:eulerarnold}) defines a family of geodesics 
$\{V^I(s)\}$ on the unitary space. The boundary condition $U(s=1)=U_{\rm target}$
filters out those geodesics that can realize the target unitary operator 
by fixing the magnitude of the tangent vector $V^I$ at $s=0$ (at the identity 
operator). This procedure is analogous to the shooting method in which the 
trajectory followed by a particle with a certain initial velocity is required 
to hit the target. The initial velocity for which the target is reached is 
determined by the boundary condition at $s=1$. 

Furthermore, there might be more than one value of the initial velocity for 
which the point of the target unitary is reached. The complexity of the 
target unitary operator is then given by the length of the shortest geodesic 
realizing the target unitary operator:
\begin{equation}
\label{eqn:complexityexpression}
	C[U_{\rm target}]= \stackunder{{\rm min}}{\{$V^I(s)$\}}\int_{0}^{1}ds\sqrt{G_{IJ}V^{I}(s)V^{J}(s)},
\end{equation}
where the minimization is over all geodesics $\{V^I(s)\}$ from the identity to $U_{\rm target}$. This equation makes use of the right-invariance of the line element (\ref{lineelement}). 

In the approximation applied here, we will keep only the leading-order term 
in the Dyson series (\ref{DysonSeries}). The implication of neglecting the higher-order 
terms in the Dyson series is that our result will correspond to upper bound on complexity 
instead of the actual complexity value. 

\begin{figure}[h!]
    \centering
    \includegraphics[scale=0.5]{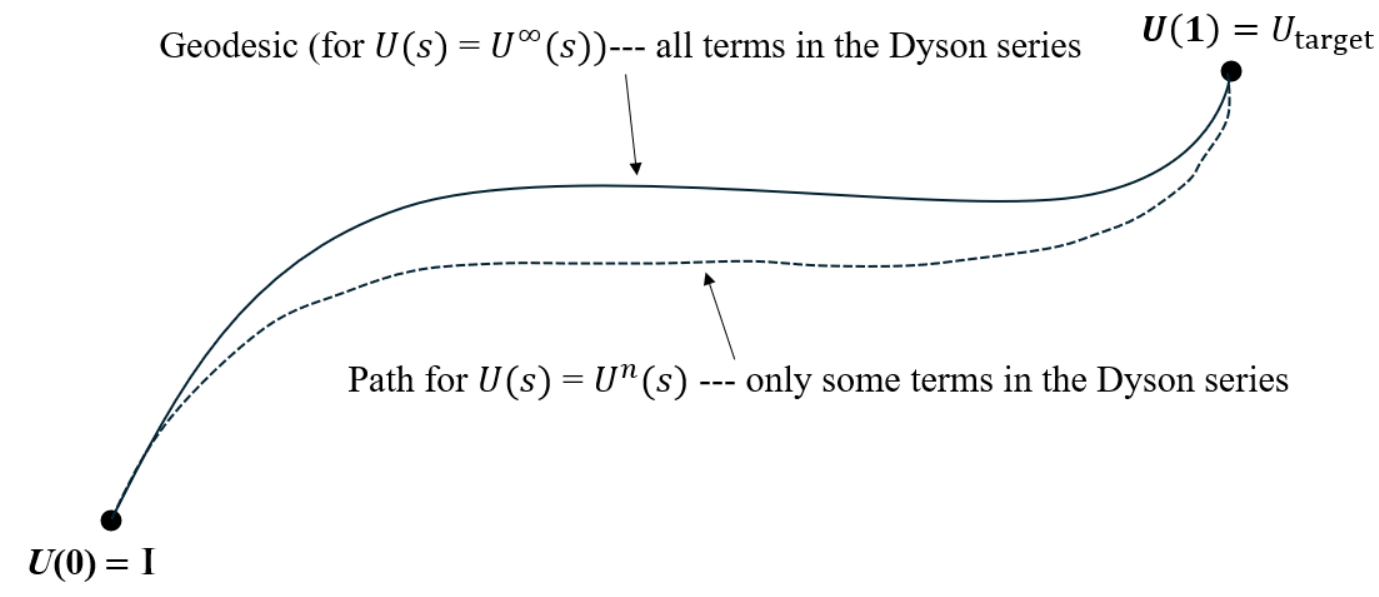}
    \caption{Pictorial interpretation of the difference between the lowest 
    complexity path and a path leading to an upper bound on the complexity. }
    \label{upperboundplot}
\end{figure}

In order to clarify it, let $U^n(s)$ represent the $n$-th order approximation to the Dyson 
series, so that $U^{\infty}(s):= \lim_{n \rightarrow \infty} U^n(s) $ corresponds to the 
geodesic curve. In consequence, by selecting the shortest path among the geodesic curves, 
the following condition holds:
\begin{align}
    C[U_{\rm target}^{\infty}]\leq C[U_{\rm target}^{n}]\ \text{for} \ n>0,
\end{align}
because, for $n\rightarrow \infty$, we obtain the exact solution(path), which is 
the shortest one, as depicted in Fig. \ref{upperboundplot}. 

One has to keep in mind that the manifold under consideration, equipped with the line 
element (\ref{lineelement}), remains unaffected by the change of $n$ and thus also 
the geodesic curve. However, by selecting a given $n$, different equations approximating 
the geodesic path on this manifold are obtained, and the exact geodesic equation is 
recovered for $U^{\infty}(s)$. 

\subsection{Recipe to determine complexity using the geometrical approach}

In this subsection, we provide a summary in the form of a simple recipe that can be applied to compute the complexity of an operator:

 \begin{itemize}
     \item From the operator whose complexity is to be determined, identify a basis 
     of the generators ($\mathcal{O}_I)$ that form a closed commutator algebra and 
     hence specify a Lie group. A commutator algebra is said to be closed if taking 
     the commutation of any two elements produces an element that is also part of 
     the algebra. It might be possible that the generators of some target operators 
     do not form a closed algebra. The simplest example is that of a unitary operator,
     which is generated from non-quadratic Hamiltonians like an anharmonic oscillator.
     In that case, by penalizing the generators of higher orders, we can neglect their
     contribution in order to regain a closed set of commutators. Geometrically, large
     penalties restrict the movement in the direction of higher-order generators.
     Alternatively, the complexity resulting from a calculation that ignores higher-order 
     generators can be interpreted as an upper bound on the exact complexity
     because it ignores possible shortcuts that might be taken in the direction of
     higher-order generators. We will also encounter additional subtleties related to 
     group-theoretical properties of non-compact Lie groups that indicate that the 
     computed distances should be considered upper bounds on the complexity rather 
     than strict values.  
     \item The commutators of the generators determine the structure constants of their
     Lie algebra. Using the structure constants, solve the Euler--Arnold equation to get 
     the set of geodesics $\{V^I(s)\}$ in the corresponding Lie group. 
     \item Having obtained the $V^I(s)$, use them to compute the path-ordered exponential 
     (\ref{sdependentunitary}).
     \item Implement the boundary conditions  $U(s=0)= \mathbb{I}$ and $U(s=1)=U_{\rm target}$ 
     by fixing the initial value of the components of the velocity vector $V^I(s)$ in terms 
     of the parameters of the target unitary. This step ensures that the geodesic characterized 
     by those initial values reaches the target.
     \item Compute the length of the geodesics for all values of the initial tangent vector 
     determined in the previous step with respect to the chosen right-invariant metric.
     \item The length of the minimal geodesic determines the complexity of the target unitary 
     operator. 
 \end{itemize}
 
It is worth repeating our new viewpoint that the geometrical method applied to an originally 
infinite-dimensional quantum system is, in general, expected to provide an upper bound on 
the complexity.  Computational tractability often requires one to ignore directions in the 
full group of unitaries that may seem unrelated to the problem at hand but might still be 
relevant for geodesics as a shortcut between the initial and target operator. We will, therefore, 
refer to our results as ``complexity bounds.''
 
We will follow this recipe for various harmonic oscillator systems, using two different 
non-compact groups that can be interpreted as containing the harmonic oscillator Hamiltonian 
as a generator. We first introduce and apply the harmonic oscillator groups, in which the 
Hamiltonian is accompanied by position, momentum, and identity, and then turns to the 
symplectic group. 

\section{The harmonic oscillator group.}
\label{sec:HOgroup}

We begin with a brief review of the harmonic oscillator group. This group is based on four 
generators $Q$, $P$, $H$ and $E$, which satisfy the commutation relations
\begin{equation}
\label{eqn:commutationHO}
    [Q,P]= i E, ~~[H,Q]=-i P,~~ [H,P]=i Q .
\end{equation}
Upon exponentiation, the generators $Q$, $P$, $H$, and $E$ specify a Lie group, which is popularly known as the \textit{harmonic oscillator group}, studied extensively in \cite{streater}. 
If we represent the generators $Q$, $P$, $H$ and $E$ as 
\begin{align}
    Q= x,~~ P= -i \frac{\partial }{\partial x},~~ H= -\frac{\partial^2}{\partial x^2}+x^2~~ {\rm and}~~ E= \mathbb{I},
\end{align}
on the Hilbert space of square-integrable functions of $x$, they can be recognized as the position, 
momentum, Hamiltonian, and the identity operator of a harmonic oscillator. The generators $P$, $Q$, 
and $E$ form a subalgebra isomorphic to the Heisenberg algebra. 

\subsection{Mathematical properties}

Exponentiation of all four generators results in the harmonic oscillator group. 
This group is not exponential, as shown in \cite{streater}, which means that there
are some group elements that cannot be written as $\exp(-i(\alpha_1 E+\alpha_2 P+\alpha_3 Q+\alpha_4 H))$ 
with real numbers $\alpha_j$. By definition, the Lie group of the corresponding 
Lie algebra is the group generated by $\exp(-i\alpha_1 E)$, $\exp(-i\alpha_2 P)$, 
$\exp(-i\alpha_3 Q)$ and $\exp(-i\alpha_4 H)$. Trying to rewrite a generic product 
of these exponentials as an exponentiated sum of the generators, using the 
Baker--Campbell--Hausdorff formula, may result in an infinite series of the generators 
that is not contained in the Lie algebra. A generic element of the Lie group is, 
therefore, not guaranteed to be the exponential of some element of the Lie algebra. 
Mathematically, this property is related to the fact that the harmonic oscillator 
group is solvable and, unlike the Heisenberg group, not nilpotent. If complexity 
calculations are performed by exponentiating all elements of the Lie algebra and
not considering products, one does not cover the full Lie group if it is not exponential.
Geodesics in the covered part of the group may then be shortened if one makes it 
possible to include also the missing directions. As already mentioned, we for this 
and other reasons consider our results to be upper bounds on the complexity rather 
than strict values.

Another subtlety is that the harmonic oscillator algebra does not permit a
finite-dimensional representation by Hermitian matrices, which is well known
from the Heisenberg subalgebra. We will therefore use a representation-
independent derivation of some relevant properties of geodesics, based on the
right-invariance of suitable line elements and derivations in the Lie algebra.
However, certain topological properties of the desired Lie group, such as 
periodic directions, cannot be captured in this way. Yet another subtlety then
appears, related to the possible existence of different covering groups that
have the same Lie algebra. We will make an attempt to highlight such features
by carefully separating derivations that can be performed at the Lie algebra
level from those that require additional properties of the group manifold.

From the commutation relation satisfied by the generators of the harmonic 
oscillator group, it is easy to see that the non-zero structure constants
$f_{IJ}^{K}$ are:
\begin{equation}
    f_{QP}^{E}= 1,~~ f_{PQ}^{E}= -1,  ~~ f_{HQ}^{P}=-1,~~ f_{QH}^{P}=1, ~~ f_{HP}^{Q}= 1, ~~ f_{PH}^{Q}= -1 .
\end{equation}
The structure constants allow us to look for directions of geodesics in 
the harmonic oscillator group, given by components of the Euler-Arnold equation 
(\ref{eqn:eulerarnold}). Choosing a diagonal penalty matrix $G_{IJ}$, these equations read
\begin{align}
\label{eqn:EulerArnoldHO}
    G_{HH}\frac{dV^H}{ds} &= (G_{QQ}-G_{PP})V^P V^Q, \\
    G_{PP}\frac{dV^P}{ds} &= -G_{QQ}V^H V^Q- G_{EE} V^Q V^E, \\
    G_{QQ}\frac{dV^Q}{ds} &= G_{PP}V^H V^P+ G_{EE} V^P V^E, \\
    G_{EE}\frac{dV^E}{ds} &= 0.
\end{align}

As a specific example of a diagonal penalty factor matrix, we implement 
equal penalties for all the generators, choosing $G_{IJ}=\delta_{IJ}$, 
such that the Euler-Arnold equations decouple:
\begin{align}
    \frac{dV^H}{ds} &=0, \\
    \frac{dV^P}{ds} &= -V^HV^Q-V^QV^E, \\
    \frac{dV^Q}{ds} &= V^HV^P+V^PV^E, \\
    \frac{dV^E}{ds} &= 0,
\end{align}
to which we have complete solutions given by: 
\begin{align}
    V^H(s) &= v_H ,~~ V^E(s) = v_E, \\
    V^P(s) &= v_P \cos (s (v_E+v_H))-v_Q \sin (s (v_E+v_H)), \\
    V^Q(s) &= v_Q \cos (s (v_E+v_H))+v_P \sin (s (v_E+v_H)).
\end{align}
More generally, to decouple the equations, it is sufficient to have 
$G_{QQ}=G_{PP}$ while $G_{HH}$ and $G_{EE}$ may be chosen independently 
(the equations do not depend on $G_{HH}$ in this case, but they do 
depend on $G_{EE}$).

Thus we have the tangent vectors $\{V^I(s)\}$ along all geodesics in 
the harmonic oscillator group manifold. The constants $v_i$ determine 
the magnitude of the velocity vectors at $s=0$, which will be fixed 
by the target unitary.  The length of the resulting curve, starting 
from the identity, equals:  
\begin{align}
    \int_0^1 \sqrt{G_{IJ}V^IV^J}ds = \sqrt{v_H^2+v_P^2+v_Q^2+v_E^2}.
\end{align}

The integrand is independent of $s$ in this case, and therefore the 
information about the path length is contained entirely in the magnitude 
of the tangent vector $V^I$ at the identity given by $v_i$. This initial 
magnitude is expressed in terms of the parameters of the target unitary 
operator. 

As a final step, we should then look at the group manifold and determine if 
topological properties imply that the same target unitary can be reached in 
multiple ways, in which case we would identify the shortest possible connection 
with the complexity bound. Such considerations cannot be performed at the Lie algebra 
level and depend on the specific choice of a target unitary.

Furthermore, let us note that the complexity bound is not 
invariant under a canonical transformation of the classical pair $(Q,P)$ 
even if it simply rescales the variables by a constant, such as 
$Q\mapsto \lambda Q$ and $P\mapsto P/\lambda$. At the quantum level, 
such a transformation redefines the penalty coefficients of $Q$ and 
$P$ and therefore changes the complexity bound.

\subsection{Computation of complexity of various target unitary operators}
\label{sec:computationcomplexity}

We now discuss the methodology of solving equation (\ref{differentialU}) 
subject to the boundary conditions (\ref{boundary}). To begin with, we should 
find a sufficiently generic expression for elements of the harmonic oscillator 
group. At this initial step, the non-exponential nature of the group is already 
relevant. If the Lie algebra has a matrix or operator representation of $N$ Hermitian 
generators $\hat{\mathcal{O}}_I$ satisfying  
\begin{equation}
\label{structureHO}
    [\hat{\mathcal{O}}_I,\hat{\mathcal{O}}_J]= i f^K_{IJ} \hat{\mathcal{O}}_{K},
\end{equation}
a large number of elements of the corresponding Lie group are obtained as
\begin{equation}
\label{factorizedU}
    U= \exp\bigg(-i \sum_I^N \alpha_I \hat{\mathcal{O}}_I\bigg)
\end{equation}
with $N$ real numbers $\alpha_I$. 
The Lie group is generated by all elements of the form $U$, taking all possible products, but it may 
contain other elements not included in the image of the exponential map. 
If the image is the whole group, the Lie group is called exponential, 
which is the case in several well-known examples, including connected 
compact Lie groups (such as ${\rm SU}(N)$) and nilpotent Lie groups 
(such as the Heisenberg group). If the Lie algebra is nilpotent, for 
instance, the Baker--Campbell--Hausdorff formula can always be used 
to bring a product of elements $U$ to the form of a single exponentiated 
sum of generators because iterating the commutator then, by definition, 
always produces zero after a finite number of iterations. The harmonic 
oscillator group, however, is not nilpotent but solvable, and it is 
not exponential, as shown in \cite{streater}. 

Nevertheless, we will use the form written in (\ref{factorizedU}), which is 
sufficient for a large set of target unitaries, although not all 
possible ones. Throughout the central part of this paper, we will assume group
elements of the form $U$ because it greatly simplifies computations. For 
comparison, we provide a detailed example using a product of exponentiated 
generators for the generic element in Appendix~\ref{appB}. In general, our 
results then provide upper bounds on the complexity but not necessarily its 
actual value because the geodesic distance through the image of the exponential 
map could be further shortened by moving through its complement in the full Lie group.

\subsubsection{Euler--Arnold equations and solutions}

{Having obtained the $V^I(s)$ for the Harmonic oscillator group by solving 
the Euler--Arnold equation, the next step is to use the obtained $V^I(s)$ 
in the Dyson series, whose leading order term can be written as: 
\begin{align*}
    -i\int_0^s V^I(s')O_I ds' &= -i\int_0^s \bigg(v_H H+ (v_P\cos(s(v_E+v_H))-v_Q \sin(s(v_E+v_H)))P \\ &~~~ + (v_Q\cos(s(v_E+v_H))+v_P \sin(s(v_E+v_H)))Q + v_E E\bigg) \\
    &= -i \bigg(\bigg\{\frac{1}{v_E+v_H}(-v_Q+v_Q \cos(s(v_E+v_H))+ v_P \sin(s(v_E+v_H))\bigg\}P \\
    & + \bigg\{\frac{1}{v_E+v_H}(v_P-v_P \cos(s(v_E+v_H))+ v_Q \sin(s(v_E+v_H)))\bigg\}Q \\ & + s v_H H + s v_E E
    \bigg).
\end{align*}

Keeping only up to the leading order term in the Dyson series, the path-ordered 
exponential can be approximately written as: 
\begin{align}
\nonumber
    U(s) &\approx \exp\bigg(-i \bigg(\bigg\{\frac{1}{v_E+v_H}(-v_Q+v_Q \cos(s(v_E+v_H))+ v_P \sin(s(v_E+v_H))\bigg\}P \\ \nonumber
    & + \bigg\{\frac{1}{v_E+v_H}(v_P-v_P \cos(s(v_E+v_H))+ v_Q \sin(s(v_E+v_H)))\bigg\}Q \\ & + s v_H H + s v_E E
    \bigg)\bigg).
\end{align}
Let us rewrite the above equation as: 
\begin{align}
\label{eqn:factorizedunitary}
    U(s) \approx \exp\bigg(-i (\alpha_1(s) E+ \alpha_2(s) P + \alpha_3(s) Q+ \alpha_4(s) H)\bigg),
\end{align}
where the $\alpha_I$ are: 
\begin{align}
    \alpha_1(s) &= s v_E,\\
	\alpha_2(s) &= \frac{1}{v_E+v_H}\bigg(-v_Q+v_Q \cos(s(v_E+v_H))+ v_P \sin(s(v_E+v_H))\bigg), \\
	\alpha_3(s) &= \frac{1}{v_E+v_H}\bigg(v_P-v_P \cos(s(v_E+v_H))+ v_Q \sin(s(v_E+v_H))\bigg), \\
	\alpha_4(s) &= s v_H.
\end{align}

In principle, the complete $s$-dependent unitary with a given tangent 
direction should be the path-ordered exponential (which involves taking 
into account all the terms in the Dyson series), and is not necessarily
equal to the simple exponential as written in equation (\ref{eqn:factorizedunitary}). 
In writing $U(s)$ as the simple exponential (by considering only the leading
order term in the Dyson series), we are implicitly assuming that the 
deviations between the simple exponential and the path-ordered one remain 
small for sufficiently short geodesic distances. As we will see, the 
expressions are equal for our leading-order results in the case of harmonic 
oscillators, but not necessarily when we introduce additional terms by 
perturbation theory.

Imposing the boundary condition at $s=1$, setting $U(s=1)= U_{\rm target}$,
allows us to determine the geodesic constants $v_I$ in terms of the target 
operator quantities. In the following sections, we consider various target 
unitary operators for the purpose of illustration and computing their 
complexities or upper bound on complexities. 

\subsubsection{Displacement operator}
\label{SectionDisplacementOperator}

We illustrate the methodology by explicitly computing the complexity 
of an operator constructed out of the harmonic oscillator group generators, 
the displacement operator
\begin{align}
    U_{\rm target}= \exp(\alpha a^{\dagger}-\alpha^{*} a),
\end{align}
for a complex number $\alpha$. The $a^{\dagger}$ and $a$ are 
creation and annihilation operators, satisfying the standard 
commutation relation $[a,a^{\dagger}] = \mathbb{I}$.

In terms of the generators of the harmonic oscillator group, this operator 
takes the form:
\begin{align}
\label{displacementoperator}
\nonumber
    U_{\rm target} &= \exp\bigg(\frac{\alpha}{\sqrt{2}} (Q-i P)-\frac{\alpha^{*}}{\sqrt{2}}(Q+iP)\bigg)  
    = \exp\bigg(\frac{1}{\sqrt{2}}(\alpha -\alpha^*) Q - \frac{i}{\sqrt{2}}(\alpha+\alpha^*)P\bigg) \\
    &= \exp\bigg(i\sqrt{2}({\rm Im}(\alpha)Q-{\rm Re}(\alpha)P)\bigg).
\end{align}
Matching the boundary conditions, we obtain: 
\begin{align}
    &\exp\bigg(-i \{\alpha_1(1) E+ \alpha_2(1) P+ \alpha_3(1) Q+ \alpha_4(1) H\}\bigg)= \exp\bigg(i\sqrt{2}({\rm Im}(\alpha)Q-{\rm Re}(\alpha)P)\bigg).
\end{align}
Comparing the coefficients of the generators on both sides, we find:
\begin{align}
    \alpha_1(1) &=0 \mbox{ and }
    \alpha_4(1) =0,  \\
     \alpha_2(1) &= \sqrt{2}{\rm Re}(\alpha) =\frac{-v_Q+v_Q \cos(v_E+v_H)+ v_P \sin(v_E+v_H)}{v_E+v_H}, \\
    \alpha_3(1) &= -\sqrt{2}{\rm Im}(\alpha) = \frac{-v_P \cos (v_E+v_H)+v_Q \sin (v_E+v_H)+v_P}{v_E+v_H}. 
\end{align}
The first equation requires $v_E=0$ and $v_H=0$, such that we have to take the appropriate limit of $v_E-v_H\to0$ when we evaluate the other two equations.
Using
\begin{align}
    \lim_{v_E+v_H \to 0} \frac{1-\cos(v_E+v_H)}{v_E+v_H}=0,
\end{align}
and
\begin{align}
    \lim_{v_E+v_H \to 0} \frac{\sin(v_E+v_H)}{v_E+v_H}= 1,
\end{align}
 we obtain: 
 \begin{align}
     v_P= - \sqrt{2} {\rm Re}(\alpha), ~~~~ {\rm and}~~~~ v_Q= \sqrt{2} {\rm Im}(\alpha).
 \end{align}

Therefore, the complexity bound of the displacement operator is given by: 
 \begin{align}
     C[D]=\sqrt{v_E^2+v_P^2+v_Q^2+v_H^2}= \sqrt{2} |\alpha|.
 \end{align}

 \subsubsection{Complexity of the time evolution operator}

In our main examples, we consider the unitary operator produced 
by exponentiation the harmonic oscillator Hamiltonian:
\begin{align}
\label{HarmonicoscillatorH}
	H_{\omega}= \frac{P^2}{2m}+\frac{m \omega^2}{2} Q^2.
\end{align}
We first assume that $m=\omega^{-1}$, which happens to simplify the 
resulting complexity bound. Classically, this relationship can always be 
achieved by a canonical transformation: $Q\mapsto \sqrt{m\omega}Q$, 
$P\mapsto P/\sqrt{m\omega}$. However, as already mentioned, such a 
transformation upon quantization changes the penalty factors and, 
therefore the complexity bound.

In terms of the generators of the harmonic oscillator group, the 
Hamiltonian $H_{\omega}$ with $m=\omega^{-1}$ can be written as:
\begin{align}
	H_{\omega}= \omega H.
\end{align}
The associated unitary operator is:
\begin{equation}
	U_{\omega}(t)=\exp\bigg(-i \omega t H\bigg).
\end{equation}
Imposing the boundary condition $U(s=1)= U_{\omega}(t)$, we have:
\begin{align}
    \exp\bigg(-i \{\alpha_1(1) E+ \alpha_2(1) P+ \alpha_3(1) Q+ \alpha_4(1) H\}\bigg)= \exp\bigg(-i \omega t H\bigg),
\end{align}
and therefore:
\begin{align}
     &v_E=0, ~~~ \frac{1}{v_E+v_H}\bigg(-v_Q+v_Q \cos(v_E+v_H)+ v_P \sin(v_E+v_H)\bigg) =0, \\
    & \frac{1}{v_E+v_H}\bigg(v_P-v_P \cos(v_E+v_H)+ v_Q \sin(v_E+v_H)\bigg)= 0 , ~~~ v_H= \omega t.
\end{align}
We obtain the solutions:
\begin{align}
\label{initialvelocities}
    v_E=0,~~~ v_P= 0, ~~~ v_Q= 0, ~~~ v_H= \omega t.
\end{align}
The corresponding length of a curve from the identity to the harmonic oscillator evolution operator with $m=\omega^{-1}$ is given by: 
\begin{align}
    L_{[U_{\rm target}]}(\omega,t)= \sqrt{v_E^2+v_P^2+v_Q^2+v_H^2}= \omega t.
\end{align}

This value may be improved further as a complexity bound because of 
possible periodic directions in the group manifold. If this length 
would always equal the complexity, the latter would grow linearly 
in time, which should not be the case according to previous derivations, such as \cite{Haque:2021hyw}. 
Periodicity properties cannot be determined solely at the Lie algebra 
level because they depend on topological properties of the Lie group 
and the specific covering space suitable for physical properties of 
the system. In the present case, we know that the harmonic oscillator 
Hamiltonian corresponding to our $H$ has spectrum $n+1/2$ with integer 
$n$. Therefore, if the finite-dimensional Lie group used here is embedded 
in the infinite-dimensional Hilbert space of quantum mechanics, 
$\exp(-i\omega tH)$ as a function of $\omega t$ has a period of $4\pi$. 
The length $L_{[U_{\rm target}]}(\omega,t)$, therefore, equals $\omega t$ 
only as long as it is less than half the period, in which case there is 
no shortcut to the same target unitary. For $2\pi<\omega t<4\pi$, however, 
the target unitary can be reached in a shorter distance by moving in 
the opposite direction, and for $\omega t=4\pi$, the target unitary 
equals the identity and has zero complexity. This process can be repeated 
for larger values of $\omega t$, resulting in the complexity bound
\begin{align}\label{Cperiodic}
    C_{[U_{\rm target}]}(\omega,t)= \sqrt{v_E^2+v_P^2+v_Q^2+v_H^2}= |\omega t-4\pi \lfloor(\omega t+2\pi)/(4\pi)\rfloor|.
\end{align}
The plot of this function is shown in Fig. \ref{fig:CompeityHO4pi}.
\begin{figure}
    \centering
    \includegraphics[scale=0.6]{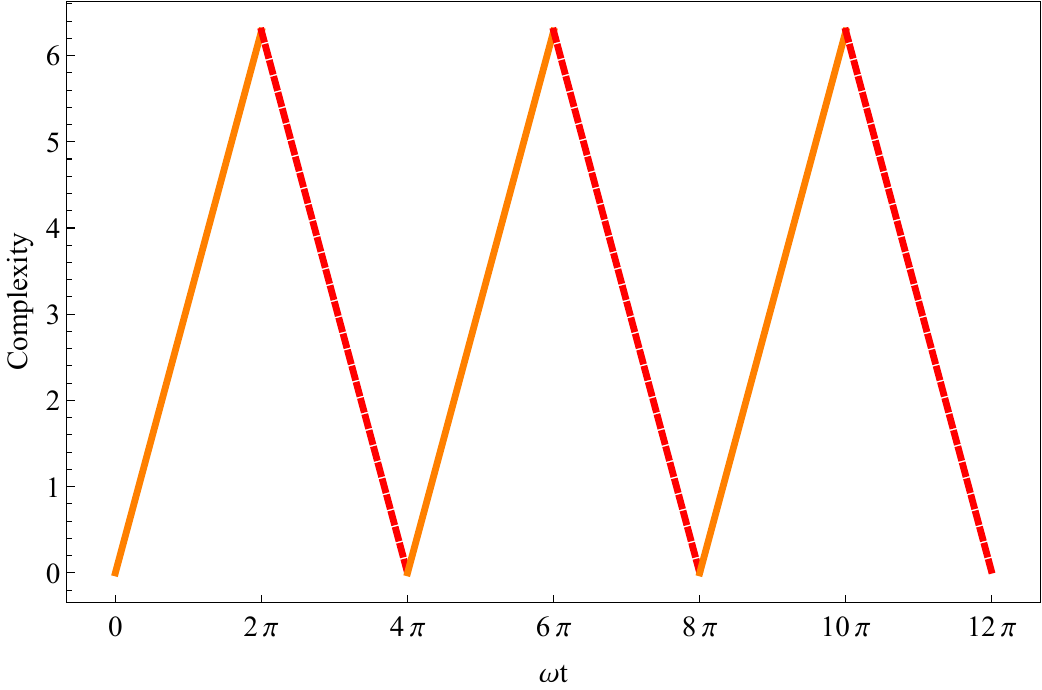}
    \caption{Complexity of the time evolution operator of a harmonic oscillator.}
    \label{fig:CompeityHO4pi}
\end{figure}

It is worth emphasizing that up to the value of the period, our result agrees with the 
complexity of the harmonic oscillator derived in Ref. \cite{Haque:2021hyw}. 
In that article, the period $T=2\pi/\omega$ of the classical 
harmonic oscillator was used. However, as indicated here, the actual 
quantum periodicity for the infinite-dimensional Hilbert space of the 
full quantum harmonic oscillator is doubling that period due to the presence 
of the $1/2$ term in its spectrum, corresponding to the non-zero ground state energy.  

Furthermore, the specific solutions of the Euler-Arnold equations 
considered here are also periodic. However, this property is independent 
of the Lie group or the embedding in the full space of unitaries,
and, therefore, cannot be considered an indicator of the periodicity 
of the complexity. 

\subsubsection{Harmonic oscillator with an additional linear potential}

As a combination of two different generators, we now  consider the time 
evolution operator of the harmonic  oscillator with an added linear potential 
as the target unitary operator, given by:
\begin{align}
    H_{\lambda}= \omega\bigg(\frac{P^2}{2}+\frac{Q^2}{2}\bigg) + \lambda Q,
\end{align}
if we still assume $m=\frac{1}{\omega}$. Classically, the added term merely
shifts the origin of the harmonic oscillator because the potential equals 
$\frac{1}{2}\omega Q^2+\lambda Q=\frac{1}{2}\omega(Q+\lambda/\omega)^2-\frac{1}
{2}\lambda^2/\omega^2$. As an expression in the harmonic oscillator algebra, 
however, the constant terms require the generator $E$, and since the product 
$EQ$ is not defined in the Lie algebra, the quadratic completion cannot be done
at this level. Within the truncated setting, the complexity bound of a harmonic 
oscillator with an additional linear term therefore need not equal the complexity bound
of an individual oscillator just computed.

The time evolution operator in this case is:
\begin{align}
    U_{\rm target}(t)= \exp(-i H_{\lambda} t)= \exp\bigg(-i\bigg\{\omega\bigg(\frac{P^2}{2}+\frac{Q^2}{2}\bigg)+\lambda Q\bigg\}t\bigg),
\end{align}
or
\begin{align}
    U_{\rm target}= \exp(-i (\omega H+\lambda Q)t),
\end{align}
in terms of the generators of the Harmonic oscillator group.
Substituting the boundary condition that $U(s=1)= U_{\rm target}$, we have: 
\begin{align}
     \exp\bigg(-i \{\alpha_1(1) E+ \alpha_2(1) P+ \alpha_3(1) Q+ \alpha_4(1) H\}\bigg)= \exp(-i (\omega H+\lambda Q)t),
\end{align}
which implies the conditions: 
\begin{align}
    &v_E=0, ~~~ \frac{1}{v_E+v_H}\bigg(-v_Q+v_Q \cos(v_E+v_H)+ v_P \sin(v_E+v_H)\bigg) =0, \\
    & \frac{1}{v_E+v_H}\bigg(v_P-v_P \cos(v_E+v_H)+ v_Q \sin(v_E+v_H)\bigg)= \lambda t , ~~~ v_H= \omega t. \label{vlambda}
\end{align}

The generator $H$ still belongs to the periodic direction in the group, 
with the period of $4\pi$ for $v_H$ if the group is to be embedded in the 
full space of unitaries. Within a period, there are now values, specifically 
given by $v_H=2\pi n$, with $n \in \mathbb{Z}$, for which the conditions have 
no solutions because the first equation in (\ref{vlambda}) then reads 
$0=\lambda t$. Here, we encounter another subtlety: A metric on a non-compact
group manifold is not guaranteed to be geodesically complete. There are then
pairs of endpoints for which the geodesic equation has no solutions, as seen 
here in a specific example. Formally, the geodesic distance between these points
is then infinite, and while it may be finite if the endpoints are moved slightly,
the divergence shows, again, that the result can only be considered an upper bound 
on the complexity. This notion of interpreting geodesic distance as the upper bound on complexity was previously done in \cite{Craps:2022ese,Craps:2023rur}.
The solutions
\begin{equation}
    v_H=|\omega t-4\pi\lfloor (\omega t+2\pi)/ (4\pi)\rfloor|
\end{equation}
\begin{equation}
v_Q=\frac{1}{2}v_H\lambda t \cot(v_H/2),
\end{equation}
and 
\begin{equation}
v_P=\frac{1}{2}v_H\lambda t,
\end{equation}
where $v_H$ is periodic as in (\ref{Cperiodic}) explicitly show the 
divergence of $v_Q$ at $v_H=2\pi n$, with $n \in \mathbb{Z}$, and 
so does the complexity bound:
\begin{equation}
    C_{[U_{\rm target}]}(\omega,\lambda,t) \leq v_H  \sqrt{1+\frac{\lambda^2t^2}{4\sin^2 (v_H /2)}},
\end{equation}
with $v_H$ as in (\ref{Cperiodic}). So that in the 
$\lambda\rightarrow 0$ limit the bound reduces to $\omega t$, 
as expected. Examples of the bound are shown in Figure~\ref{figlinearnew}. 

\begin{figure}[h!]
    \centering
    \includegraphics[scale=0.6]{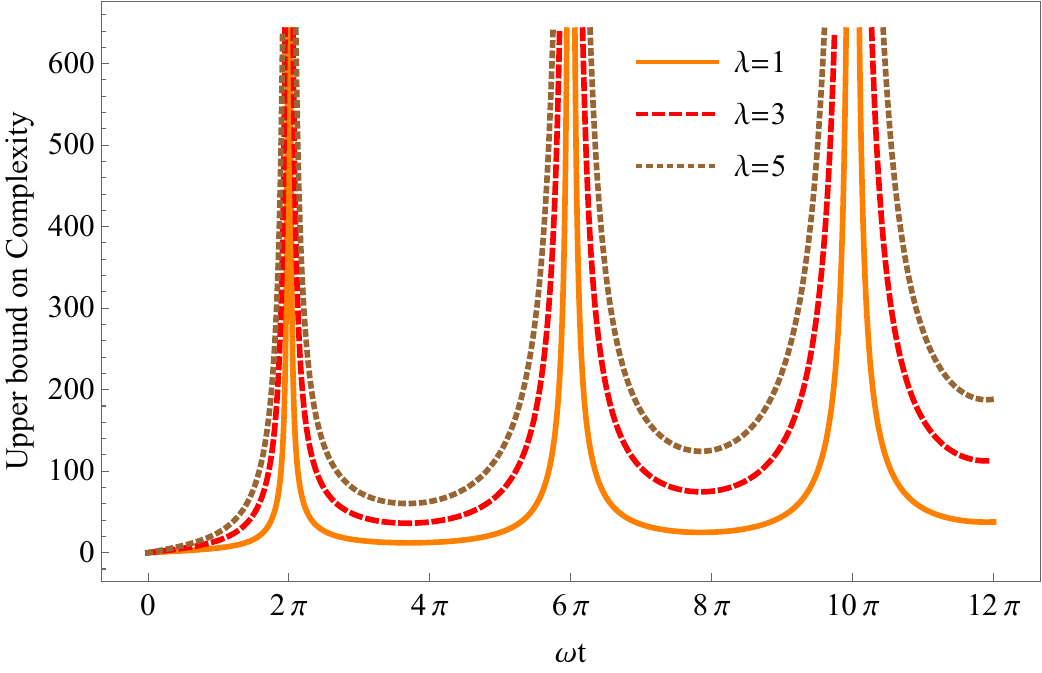}
    \caption{Complexity bound of the time evolution operator of a harmonic oscillator with an additional linear term in Q.}
    \label{figlinearnew}
\end{figure}
As a result, introducing an extra direction associated with the generator $Q$ 
can potentially lead to increased complexity, possibly even approaching infinity.
From a mathematical perspective, such a scenario cannot be excluded when dealing with non-compact groups. Nevertheless, 
it is important to note that infinite complexity values are not expected to arise 
in physical situations. Such extreme values would violate
\emph{Lloyd’s computational bound} \cite{Lloydbound, Brown:2015bva}, making them 
highly unlikely to be encountered. In our context, non-compact groups can lead to geodesic incompleteness such that there may be no finite-length geodesic between a given pair of two points. Interpreting our results as upper bounds on the complexity rather than precise values, which is also indicated by the truncations we use of the original infinite-dimensional group of unitary operators as well as the non-exponential nature of the groups used here, is consistent with both the mathematical and physical viewpoints. This question will be discussed in more detail in our conclusions.

\subsubsection{Geodesic equation for general frequencies}

So far, we have restricted our attention to harmonic oscillators with the 
condition $m=\omega^{-1}$. At the quantum level, as already noted, it is 
not possible to use a canonical transformation in order to bring any harmonic 
oscillator to this form because this step would modify the penalty factors. 
Alternatively, the harmonic oscillator algebra can be generalized to include 
an arbitrary Hamiltonian with a generic frequency $\omega$ as well as an 
independent mass $m$ by using: 
\begin{equation}
    H=\frac{1}{2}\frac{P^2}{m}+\frac{1}{2}m\omega^2Q^2,
\end{equation}
as a generator, together, with the previous $Q$, $P$, and $E$. The structure 
constants are now modified because the relevant commutation relations are: 
$[H,Q]=-iP/m$ and $[H,P]=im\omega^2Q$. 

As a consequence, the Euler--Arnold equations take the modified form:
\begin{align}
    G_{HH}\frac{dV^H}{ds} &= (m\omega^2G_{QQ}-m^{-1}G_{PP})V^P V^Q, \\
    G_{PP}\frac{dV^P}{ds} &= -m\omega^2G_{QQ}V^H V^Q+ G_{EE} V^Q V^E, \\
    G_{QQ}\frac{dV^Q}{ds} &= m^{-1}G_{PP}V^H V^P- G_{EE} V^P V^E, \\
    G_{EE}\frac{dV^E}{ds} &= 0.
\end{align}
If we continue to use the penalty factor matrix $G_{IJ}=\delta_{IJ}$, 
these equations do not decouple, and they are non-linear and much harder 
to solve. 

Another possibility to change the frequency is to add a term quadratic 
in $Q$ to the harmonic oscillator Hamiltonian, instead of a linear term
as in the preceding example. This term is not contained in the harmonic 
oscillator algebra, but it can be included in a ${\rm sp}(2,{\mathbb R})$ algebra, 
which we will describe in detail in the next section.

\section{Harmonic oscillator complexity from the ${\rm sp}(2,{\mathbb R})$ Lie algebra}
\label{sec:complexitySU}

The quadratic nature of the harmonic oscillator Hamiltonian makes it 
possible to view it as a generator of various Lie algebras. Let us 
again consider the Hamiltonian:
\begin{align}
     \hat{H}= \frac{P^2}{2 m}+\frac{1}{2}m \omega^2 Q^2,
     \label{HOHamiltMass}
\end{align}
with independent mass and frequency.

A natural set of operators that form a closed Lie algebra 
includes $\frac{1}{2}\hat{P}^2$, $\frac{1}{2}\hat{Q}^2$ 
and their commutator. This observation suggests using the 
generators:
\begin{align}
    K_{1} &= \frac{\hat{Q}^2}{2}, \\
    K_{2} &= \frac{\hat{P}^2}{2}, \\
    K_{3} &= \frac{1}{2}(\hat{Q}\hat{P}+\hat{P}\hat{Q}),
\end{align}
which satisfy the following commutation relations:
\begin{align}
    [K_{1},K_{2}]= i K_3 ~~~~ [K_3, K_{1}]= -2 i K_1,~~~ [K_3, K_{2}]=2 i K_2.
    \label{su11algebra}
\end{align}

Since these three operators form a closed Lie algebra, given by 
${\rm sp}(2,{\mathbb R})$, they specify a Lie group, ${\rm Sp}(2,{\mathbb R})$ or one of its covering groups, part of which can be parameterized by the group elements
\begin{align}
    U(s)= \exp(-i (\gamma_1(s) K_1 +\gamma_2(s) K_2+ \gamma_3(s) K_3)).
\end{align}
(The Lie group ${\rm SU}(1,1)$, which has been used in the past for this system, is a different real form.) According to \cite{exp}, the Lie group ${\rm Sp}(2,{\mathbb R})$ is not exponential, 
but the projective version, isomorphic to ${\rm SO}(2,1)$ of which ${\rm Sp}(2,{\mathbb R})$ is a 2-fold covering, is. A large part of ${\rm Sp}(2,{\mathbb R})$, but not all of it, can therefore be parameterized by exponentiated sums of the generators.

The structure constants associated with the Lie algebra (\ref{su11algebra}) are:
\begin{align}
    f_{12}^3= 1 , ~~~ f_{31}^{1}=-2 ,~~~ f_{32}^{2}= 2.
\end{align}
In consequence, the Euler--Arnold equations (choosing $G_{IJ}=\delta_{IJ}$ 
as before) can be written as: 
\begin{align}
    \frac{dV^1}{ds} &= V^2 V^3+2 V^1 V^3,\\
    \frac{dV^2}{ds} &= -V^1 V^3-2 V^2 V^3, \\
    \frac{dV^3}{ds} &= -2(V^1)^2+2(V^2)^2.
\end{align}
The system at hand consists of three coupled non-linear 
differential equations. We can significantly simplify these 
equations by choosing an alternative basis of the Lie algebra, 
which is given by:
\begin{align}
 J_1 &= K_3 = \frac{1}{2}(QP+PQ), \\
 J_2 &= K_1-K_2 = \frac{1}{2}(Q^2-P^2), \\
 J_3 &= K_1 + K_2 = \frac{1}{2}(Q^2+P^2), 
\end{align}
with commutation relations:
\begin{align}
    [J_1,J_2]= -2 i J_3, ~~~[J_2, J_3]= 2 i J_1, ~~~[J_3,J_1]= 2 i J_2.
\end{align}
In this form, the Euler-Arnold equations (still with $G_{IJ}=\delta_{IJ}$) 
reduce to: 
\begin{align}
    \frac{dV^1}{ds} &= -4 V^2V^3,\\
    \frac{dV^2}{ds} &= 4 V^1 V^3,\\
    \frac{dV^3}{ds} &= 0,
\end{align}
for which the solutions are given by:
\begin{align}
    V^1(s) &= v_1 \cos(4 s v_3)- v_2 \sin(4 s v_3), \\
    V^2(s) &= v_2 \cos(4 s v_3)+ v_1 \sin(4 s v_3), \\
    V^3(s) &=  v_3.
\end{align}

Keeping up to leading-order terms in the Dyson series, the path-ordered 
exponential can be approximately written as: 
\begin{align}
\label{genericSU}
    U(s) \approx \exp(-i(\gamma_1(s) J_1 + \gamma_2(s) J_2+ \gamma_3(s) J_3)),
\end{align}
where:
\begin{align}
    \gamma_1(s) &= \frac{1}{4 v_3}\bigg(-v_2+v_2 \cos(4 s v_3)+ v_1 \sin(4 s v_3) \bigg), \label{solgamma1}\\
    \gamma_2(s) &= \frac{1}{4 v_3}\bigg(v_1-v_1 \cos(4 s v_3)+ v_2 \sin(4 s v_3) \bigg), \label{solgamma2} \\
    \gamma_3(s) &= s v_3. \label{solgamma3}
\end{align}

\subsection{Time evolution operator of the harmonic oscillator -- revisited}

Employing the ${\rm sp}(2,{\mathbb R})$ algebra, let us now revise the 
geometric complexity of the time evolution operator generated 
by the harmonic  oscillator Hamiltonian (\ref{HOHamiltMass}).
Again, we first choose $m=\omega^{-1}$, so that:
\begin{align}
    \mathbf{H}= \omega\bigg(\frac{Q^2}{2}+\frac{P^2}{2}\bigg) = \omega J_3.
\end{align}
In consequence, the target unitary operator is:
\begin{align}
    U(s=1)=U_{\rm target} = \exp(-i \omega t J_3),
\end{align}
where $U(s=1)$ can be found by substituting the solutions (\ref{solgamma3}) in (\ref{genericSU}):
\begin{align}
    \exp(-i(\gamma_1(1)J_1+ \gamma_2(1) J_2+ \gamma_3(1) J_3))= \exp(-i \omega t J_3).
\end{align}
which implies the conditions: 
\begin{align}
    \gamma_1(1)= 0, ~~~ \gamma_2(1)=0, ~~~~ \gamma_3(1) = \omega t.
\end{align}
Employing Eqs. (\ref{solgamma1},\ref{solgamma2},\ref{solgamma3}) this leads to:
\begin{align}
    v_3= \omega t,~~~ v_1=0, ~~~ v_2=0.
\end{align}

The final step requires consideration of periodicity 
properties in possible exponentiations of the Lie 
algebra ${\rm sp}(2,{\mathbb R})$. The immediate choice, ${\rm Sp}(2,{\mathbb R})$, would represent the generator $J_3$ by the matrix:
\begin{align}
 J_3= \begin{pmatrix}
     1 && 0 \\
     0 && -1
 \end{pmatrix},  
\end{align}
such that 
\begin{align}
    \exp(-i\omega t J_3)= \begin{pmatrix}
        e^{-i \omega t} && 0 \\
        0 && e^{i \omega t}
    \end{pmatrix},
\end{align}
as also used in \cite{Haque:2021hyw}. However, the resulting period of $2\pi$ in $\omega t$ obtained for $\exp(-i\omega t H)$ is not compatible with the spectrum $n+1/2$ of $\hat{H}$ in the full infinite-dimensional space of unitaries. As an abstract group, ${\rm Sp}(2,{\mathbb R})$ has infinitely many covering groups, including a 2-fold covering which results in a compatible period of $4\pi$ in $\omega t$. This covering group is the metaplectic group ${\rm Mp}(2,{\mathbb R})$, but it is not a matrix group and does not have finite-dimensional matrix representations. Therefore, it is not possible to obtain the period of $4\pi$ from a matrix calculation, following the methods of \cite{Haque:2021hyw}, but it is consistent within our treatment based mainly on the Lie algebra plus a final periodicity condition that requires only abstract group properties but no matrix representation. The resulting complexity bound is the same as in (\ref{Cperiodic}),
\begin{align}
    C_{[U_{\rm H.O.}]}(\omega,t)= \sqrt{v_E^2+v_P^2+v_Q^2+v_H^2}= |\omega t-4\pi \lfloor(\omega t+2\pi)/(4\pi)\rfloor|,
    \end{align}
with a period of $4\pi$ in $\omega t$.

\subsection{Time evolution operator of the inverted harmonic oscillator}

The same method allows us to consider the time evolution operator of the inverted harmonic oscillator as the target unitary, 
\begin{align}
    \mathbf{H_{IHO}}= \frac{P^2}{2m}-\frac{1}{2}m \Omega^2 Q^2.
\end{align}
This model was also considered in \cite{Bhattacharyya:2020art,Bhattacharyya:2021cwf} from the perspective of complexity.
Again for the sake of simplicity, we assume $m=\Omega^{-1}$. In that case the Hamiltonian in terms of the generators of the group ${\rm Sp}(2,{\mathbb R})$ can be written as 
\begin{align}
    \mathbf{H_{IHO}}= \frac{\Omega}{2}(P^2-Q^2)= -\Omega J_2,
\end{align}
such that
\begin{align}
    U_{\rm target}=U(s=1)= \mathbf{U_{IHO}}=\exp\bigg(i \Omega J_2\bigg).
\end{align}

The equation
\begin{align}
    \exp(-i(\gamma_1(1)J_1+ \gamma_2(1) J_2+ \gamma_3(1) J_3))= \exp\bigg(i \Omega tJ_2\bigg),
\end{align}
implies
\begin{align}
    \gamma_1(1)= 0, ~~~ \gamma_2(1)= - \Omega t, ~~~ \gamma_3(1)= 0.
\end{align}
Using equation~(\ref{solgamma3}), we obtain 
\begin{align}
    v_3=0,~~~ v_2= -\Omega t, ~~~ v_1 = 0,
\end{align}
with
\begin{align}
\lim_{v_3 \to 0} \frac{1-\cos(4 v_3)}{4 v_3}=0, ~ {\rm and}~ \lim_{v_3 \to 0} \frac{\sin(4 v_3)}{4 v_3}=1.
\end{align}
The resulting complexity bound equals: 
\begin{align} \label{CInv}
    C[U_{\rm I.H.O}] \leq \sqrt{v_1^2+v_2^2+v_3^2}= |\Omega t|.
\end{align}
Since $J_2$ does not correspond to a periodic direction in ${\rm Sp}(2,{\mathbb R})$ or its covering groups, there is no periodicity condition on the complexity, which instead grows linearly in time.

 \subsection{Harmonic oscillator with an additional $Q^2$ term in the Hamiltonian.}

Within ${\rm Sp}(2,{\mathbb R})$, we can change the frequency of the harmonic oscillator Hamiltonian by adding an additional quadratic potential, which in contrast to the harmonic oscillator group is now one of the generators. If we still assume $m=\omega^{-1}$ for the original Hamiltonian, adding a quadratic term with a free coefficient, as in
 \begin{align}
     \mathbf{H_{H.O.\lambda}}= \omega\bigg(\frac{P^2}{2}+\frac{1}{2}Q^2\bigg) + \lambda Q^2,
 \end{align}
allows us to interpret this system as having independent mass $m=\omega^{-1}$ and frequency: 
\begin{equation}
    \bar{\omega}=\sqrt{\frac{\omega+2\lambda}{m}}=\omega\sqrt{1+2\frac{\lambda}{\omega}}.
\end{equation}
    
 In terms of the generators $J_i$, the new Hamiltonian can be written as 
 \begin{align} \label{HJ}
     \mathbf{H_{H.O.\lambda}}= \omega J_3 + \lambda (J_2+J_3) = (\omega+\lambda)J_3 + \lambda J_2. 
 \end{align}
 Therefore, the target unitary operator is:
 \begin{align}
     U_{\rm target}= \exp(-i\{ (\omega+\lambda)J_3+ \lambda J_2\}t).
 \end{align}
Identifying the two expressions:
\begin{align}
    \exp(-i(\gamma_1(1)J_1+ \gamma_2(1) J_2+ \gamma_3(1) J_3))= \exp(-i\{ (\omega+\lambda)J_3+ \lambda J_2\}t),
\end{align}
implies the boundary conditions:
\begin{align}
    \gamma_1(1)= 0, ~~~ \gamma_2(1)= \lambda t , ~~~~ \gamma_3(1)= (\omega+\lambda) t,
\end{align}
which lead to: 
\begin{align}
\label{conditiongamma}
    \gamma_1(1) &= \frac{1}{4 v_3}\bigg(-v_2+v_2 \cos(4 v_3)+ v_1 \sin(4 v_3) \bigg)= 0, \\
    \gamma_2(1) &= \frac{1}{4 v_3}\bigg(v_1-v_1 \cos(4 v_3)+ v_2 \sin(4  v_3) \bigg)= \lambda t, \\
    \gamma_3(1) &=  v_3 = (\omega+\lambda) t.
\end{align}

The periodicity argument is now more complicated for general $\lambda$ because a full representation of the Hamiltonian operator on the infinite-dimensional Hilbert space of quantum mechanics shows that $U_{\rm target}$ should have period $4\pi$ in $\bar{\omega}t$. However, the linear combination (\ref{HJ}) in terms of ${\rm sp}(2,{\mathbb R})$ generators does not exponentiate to a periodic expression because it contains $J_2$, and the coefficient of $J_3$ which does belong to a periodic direction in the group has coefficient $\omega+\lambda\not=\bar{\omega}$. A reliable periodicity argument can be given only approximately for small $\lambda$, in which case the $J_2$-contribution to the Hamiltonian has a small coefficient, and $\omega+\lambda\approx\bar{\omega}$. Periodicity in the $J_3$-direction then approximates the period of $4\pi$ in the full representation, and we can solve the equations by:
\begin{equation}
    v_1=2v_3\lambda t,\quad\mbox{and}\quad v_2=2v_3\lambda t\cot(2v_3),
\end{equation}
where:
\begin{equation} \label{v3lambda}
 v_3=|(\omega+\lambda)t-4\pi\lfloor ((\omega+\lambda)t+2\pi)/ (4\pi)\rfloor|,
\end{equation}
for minimized geodesics. The complexity bound is the given by:
\begin{equation} \label{Clambda}
    C_{[U_{\rm target}]}(\omega,\lambda,t) \leq \sqrt{v_1^2+v_2^2+v_3^2}= v_3\sqrt{1+\frac{4\lambda^2t^2}{\sin^2(2v_3)}}\,.
\end{equation}


\begin{figure}
    \centering
    \includegraphics[scale=0.6]{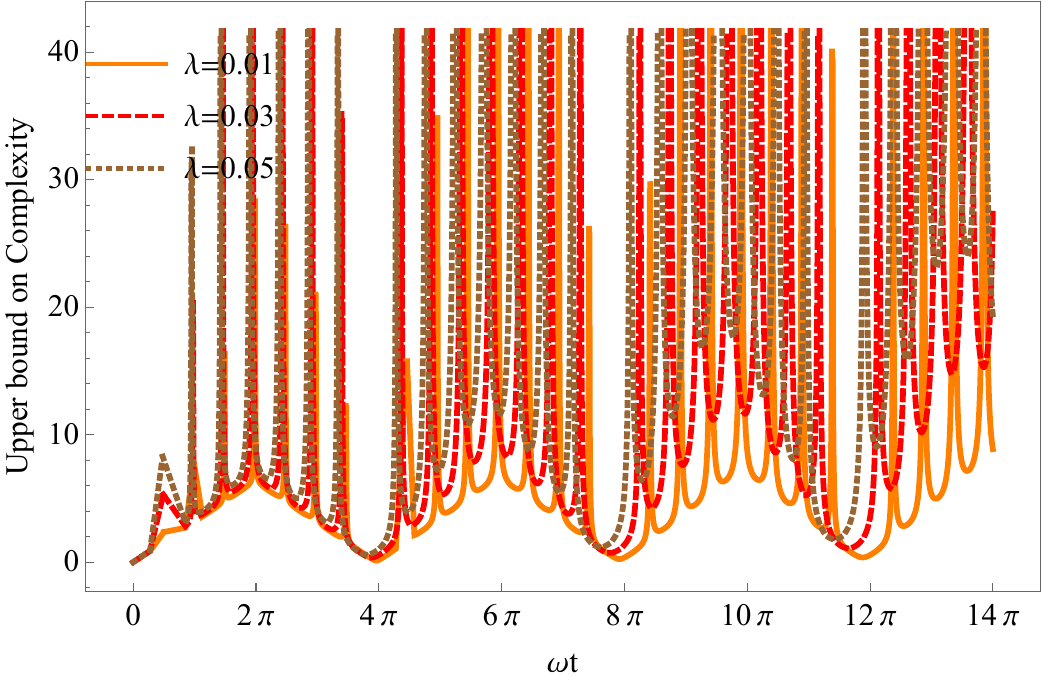}
    \caption{Complexity bounds on time evolution of a harmonic oscillator with quadratic term in Q.}
    \label{fig2}
\end{figure}

Since $\sqrt{1+2\lambda/\omega}<1+\lambda/\omega$ for $\lambda\not=0$, the complexity 
bound in the first branch of $|\bar{\omega}t|<2\pi$ is greater than $\bar{\omega}t$. 
The fact that we are using two generators, $H$ and $Q^2$, in order to construct the 
Hamiltonian means that both penalty factors are relevant. The complexity bound is, 
therefore, greater than expected for an individual oscillator of frequency $m=\omega^{-1}$, 
for which the additional $Q$-term is not needed. 

Going beyond the range of small $\lambda$, we can consider two instructive special 
cases. First, applying our result to $\lambda=-\omega/2$, we have the Hamiltonian 
for a free particle of mass $m=\omega^{-1}$. The complexity bound in this case is: 
\begin{align}
    C_{[U_{\rm target}]}(m^{-1},-(2m)^{-1},t) \leq  \sqrt{v_1^2+v_2^2+v_3^2}= v_3\sqrt{1+\frac{t^2}{m^2\sin^2(2v_3)}},
\end{align}
where 
\begin{equation}
    v_3=\frac{1}{2}|\omega t-8\pi\lfloor(\omega t+4\pi)/(8\pi)\rfloor|.
\end{equation}
In the limit of $\lambda\to-\omega$, the Hamiltonian describes the inverted Harmonic oscillator. 
Equation (\ref{v3lambda}) then implies that $v_3\to0$, which in (\ref{Clambda}) results in the complexity bound:
\begin{equation}
    C_{[U_{\rm target}]}(\omega,-\omega,t) \leq |\omega t|,
\end{equation}
consistent with our previous result for this system, equation (\ref{CInv}). 
In this limit, the periodicity argument simplifies because $J_3$ disappears 
from the Hamiltonian (\ref{HJ}), and only the non-compact $J_2$ remains. 
The complexity bound is, therefore, reliable even though $\lambda=-\omega$ 
is not small.

\section{Coupled harmonic oscillators}
\label{sec:CoupledHO}

A model of two coupled oscillators was considered in \cite{Jefferson:2017sdb}, 
but with a coupling term different from what we consider here. More importantly, 
the previous paper was interested in the complexity of the ground state of this
Hamiltonian, as briefly reviewed here. The authors of \cite{Jefferson:2017sdb} 
were interested in the complexity of:
\begin{align}
    \Psi_T = \frac{(\omega_1\omega_2-\beta^2)^{1/4}}{\sqrt{\pi}}\exp\bigg(-\frac{\omega_1}{2}x_1^2-\frac{\omega_2}{2}x_2^2-\beta x_1x_2\bigg)\,,
\end{align}
referred to as the target state, relative to the reference state: 
\begin{align}
    \Psi_R = \sqrt{\frac{\omega_0}{\pi}}\exp\bigg(-\frac{\omega_0}{2}(x_1^2+x_2^2)\bigg)\,.
\end{align}

Once the target and the reference states are fixed, the next step involves 
identifying simple gates used to construct the unitary (or quantum circuit) 
that can implement the transformation $\Psi_T= U\Psi_R$. They chose the following 
simple gates:
\begin{align}
    H= e^{i\epsilon x_0 p_0}, ~~~ J_a = e^{i \epsilon x_0 p_a}, ~~~ K_a = e^{i \epsilon x_a p_0}, ~~~ Q_{ab}= e^{i \epsilon x_a p_b}, ~~~ Q_{aa}= e^{\frac{\epsilon}{2}}e^{i\epsilon x_a p_a}\,.
\end{align}
Out of the chosen gates, it was realized that $Q_{ab}$ and $Q_{aa}$ would suffice 
for their purpose, considering the target state they were interested in, acting on 
a generic wave function as follows:
\begin{align}
    Q_{11}\Psi(x_1,x_2)= e^{\epsilon/2}\Psi(e^{\epsilon}x_1,x_2), ~~~~ Q_{21}\Psi(x_1,x_2)= \Psi(x_1+\epsilon x_2,x_2)\,.
\end{align}

These two gates played a crucial role in the circuits they constructed. 
However, finding the optimal circuit is necessary to determine the complexity. 
Nielsen's geometric approach was then utilized to complete this step. 
To apply Nielsen's geometric approach, it is necessary to understand that 
both the target and the reference states are Gaussian, such that Gaussian
wave functions represent the endpoints of any curve connecting the two states. 
Furthermore, the scaling and entangling gate actions preserve the Gaussian 
structure of the wave functions. Thus, it was concluded that the circuit 
constructed out of the $Q$ gates forms a representation of ${\rm GL}(2,\mathbb{R})$. 
Hence, finding the optimal circuit required finding the shortest geodesic 
in the space of ${\rm GL}(2,\mathbb{R})$ transformations. Of course, there is some 
arbitrariness in the obtained complexity, as it is dependent on the 
choice of the reference state and the choice of gates. 

As a first step towards understanding the quantum complexity of the time 
evolution operator of a simple interacting system in a state-independent 
way, we consider the case of two coupled harmonic oscillators with 
time-independent coupling. Although the model considered here is quite 
similar to what was considered in \cite{Jefferson:2017sdb}, our interest 
lies in the complexity of the time evolution operator. Our results then 
tell us how the complexity changes with time as the system evolves.
As the Hamiltonian of such a system, we choose:
\begin{align}
    \mathbf{H_{C.O}}= \frac{P_1^2}{2m_1}+\frac{P_2^2}{2m_2}+ \frac{1}{2m_1} \omega_1^2 Q_1^2 +\frac{1}{2m_2} \omega_2^2 Q_1^2 + \mu^2 (Q_1 Q_2+P_1P_2).
\end{align}

For the sake of simplicity, we again assume $\omega_i= m_i^{-1}$, for $i=1,2$, 
which simplifies the Hamiltonian to: 
\begin{align}
    \mathbf{H_{C.O}}= \omega_1 \bigg(\frac{P_1^2}{2}+\frac{Q_1^2}{2}\bigg)+ \omega_2 \bigg(\frac{P_2^2}{2}+\frac{Q_2^2}{2}\bigg)+ \mu^2 (Q_1Q_2+ P_1P_2),
\end{align}
and defines the target unitary operator:
\begin{align}
\label{targetcoupled}
    U_{\rm target}= \exp(-i \mathbf{H_{C.O}}t).
\end{align}

A natural choice of generators that form a closed Lie algebra for the target 
unitary operator (\ref{targetcoupled}) is $T_1=H_1= \frac{1}{2}(Q_1^2+P_1^2)$, 
$T_2=H_2=\frac{1}{2}(Q_2^2+P_2^2)$, $T_3=\frac{1}{2}(Q_1^2-P_1^2)$, 
$T_4=\frac{1}{2}(Q_2^2-P_2^2)$, $T_5=(Q_1P_1+P_1Q_1)$, $T_6=(Q_2P_2+P_2Q_2)$, 
$T_7=(Q_1Q_2+P_1P_2)$, $T_8=(Q_1P_2+P_1Q_2)$   
$T_{9}=(Q_1Q_2-P_1P_2)$, $T_{10}=(Q_1P_2-P_1Q_2)$. 
The generators form a ten-dimensional Lie algebra isomorphic to ${\rm sp}(4,\mathbb{R})$, which can be derived in the same way as used for second-order central moments in \cite{Bosonize}. For our target unitary operator, it is sufficient to consider the subalgebra formed by the four generators $T_1$, $T_2$, $T_7$ and $T_{10}$, which we relabel as $M_i$:
\begin{align}
    M_1= H_1, ~~~~ M_2 = H_2, ~~~~ M_3= Q_1Q_2+P_1P_2, ~~~~ M_4= P_1Q_2-Q_1P_2 .
\end{align}

Their commutation relations are: 
\begin{align}
    & [M_1,M_2]= 0, ~~~~ [M_1,M_3]= -i M_4 , ~~~~ [M_1,M_4]= i M_3 , ~~~ \\
    & [M_2,M_3]= i M_4 , ~~~~ [M_2,M_4]= -i M_3 , ~~~~ [M_3,M_4]= -2 i M_1+ 2 i M_2.
\end{align}
It follows that $M_1+M_2$ commutes with all other generators in this subalgebra, while $M_1-M_2$ together with $M_3$ and $M_4$ obey the commutation relations of ${\rm su}(2)$:
\begin{equation}
    [M_1-M_2,M_3]=-2iM_4,\quad [M_1-M_2,M_4]=2iM_3,\quad[M_3,M_4]=-2i(M_1-M_2)\,.
\end{equation}

The Lie algebra is, therefore, a direct product of the Abelian ${\rm u}(1)$ 
with the ${\rm su}(2)$-subalgebra of ${\rm sp}(4,\mathbb{R})$. Compactness 
and periodicity properties of possible exponentiations are simpler for this 
algebra than in the case of ${\rm sp}(2,\mathbb{R})$ or the full ${\rm sp}(4,\mathbb{R})$. 
(The universal covering group of ${\rm sp}(4,\mathbb{R})$ has been studied, 
for instance in \cite{Universal}.) In particular, the Abelian generator 
$M_1+M_2=H_1+H_2$, which has an integer spectrum in quantum mechanics, 
should exponentiate to an operator with period $2\pi$, and $M_1-M_2$ has 
the same period.

In terms of the new generators, the target unitary operator is given by: 
\begin{align}
    U_{\rm target}= \exp\bigg(-i(\omega_1 M_1+ \omega_2 M_2 + \mu^2 M_3) t \bigg).
\end{align} 
The Euler-Arnold equations are:
\begin{align}
    \frac{dV^1}{ds} &= f_{13}^4 V^3 \frac{G_{44}}{G_{11}} V^4+ f_{14}^3 V^4 \frac{G_{33}}{G_{11}}V^3 = -\frac{G_{44}}{G_{11}}V^3 V^4+ \frac{G_{33}}{G_{11}}V^3V^4, \\
    \frac{dV^2}{ds} &= f_{23}^4 V^3 \frac{G_{44}}{G_{22}} V^4+ f_{24}^3 V^4 \frac{G_{33}}{G_{22}}V^3 = \frac{G_{44}}{G_{22}}V^3 V^4- \frac{G_{33}}{G_{22}} V^3 V^4,  \\
    \frac{dV^3}{ds} &= f_{31}^4 V^1 \frac{G_{44}}{G_{33}} V^4+ f_{32}^4 V^2 \frac{G_{44}}{G_{33}}V^4 + f_{34}^1 V^4 \frac{G_{11}}{G_{33}}V^1 + f_{34}^2 V^4 \frac{G_{22}}{G_{33}}V^2 \nonumber \\
    & = \frac{G_{44}}{G_{33}}V^1 V^4- \frac{G_{44}}{G_{33}}V^2 V^4 -2 \frac{G_{11}}{G_{33}}V^4 V^1 + 2 \frac{G_{22}}{G_{33}}V^4 V^2, \\
    \frac{dV^4}{ds} &= f_{41}^3 V^1 \frac{G_{33}}{G_{44}} V^3+ f_{42}^3 V^2 \frac{G_{33}}{G_{44}}V^3 + f_{43}^1 V^3 \frac{G_{11}}{G_{44}}V^1 + f_{43}^2 V^3 \frac{G_{22}}{G_{44}}V^2 \nonumber \\
    & = -\frac{G_{33}}{G_{44}}V^1 V^3+ \frac{G_{33}}{G_{44}}V^2 V^3 +2 \frac{G_{11}}{G_{44}}V^3 V^1 - 2 \frac{G_{22}}{G_{44}}V^3 V^2.
\end{align}

As before, we will consider the case of all generators with equal penalties, 
$G_{IJ}= \delta_{IJ}$. Then the Euler-Arnold equations reduce to:
\begin{align}
     \frac{dV^1}{ds} &= 0,\\
    \frac{dV^2}{ds} &= 0,\\
    \frac{dV^3}{ds} & = -V^1 V^4 + V^2 V^4, \\
    \frac{dV^4}{ds} &= V^1V^3-V^2V^3.
\end{align}

The equations can easily be solved by:
\begin{align}
    &V^1(s)= v_1 , ~~~  V^2(s) = v_2 , ~~~ V^4(s)= v_4 \cos(s(v_1-v_2))+v_3 \sin(s(v_1-v_2)),  \nonumber \\
    & V^3(s)= v_3 \cos(s(v_1-v_2))-v_4 \sin(s(v_1-v_2)),
\end{align}
and the length of the geodesics characterized by the $v_i$ equals:
\begin{align}
    \int_0^1 \sqrt{G_{IJ}V^I V^J}ds = \sqrt{v_1^2+v_2^2+v_3^2+v_4^2}.
\end{align}

Another interesting and physically justifiable case of the penalty factor 
matrix is motivated by the specific setting of two coupled harmonic oscillators. 
Out of the four generators, $M_1$ and $M_2$ consist of terms associated 
with only one individual oscillator, while the generators $M_3$ and $M_4$ 
involve terms that act on both oscillators. It is natural to assign higher 
penalties to generators that involve two oscillator terms compared to the 
terms that act on only one of them. We may, therefore, choose: 
\begin{align}
    G_{11}= G_{22} := q,~~~~  \text{and}~~~~ G_{33}= G_{44} :=p ~~~~~~ (p>q).
\end{align}

With this choice, the Euler--Arnold equations are given by: 
    \begin{align}
    \frac{dV^1}{ds} &= 0,\\
    \frac{dV^2}{ds} &= 0,  \\
    \frac{dV^3}{ds} &= \bigg(1-\frac{2q}{p}\bigg) V^1V^4- \bigg(1-\frac{2q}{p}\bigg) V^2V^4,\\
    \frac{dV^4}{ds} &= -\bigg(1- \frac{2q}{p} \bigg)V^1 V^3+ \bigg(1- \frac{2q}{p} \bigg)V^2 V^3 ,
\end{align}
and can be solved as: 
\begin{align}
\label{soleulerarnold}
    & V^1(s)=v_1, \\
    & V^2(s)= v_2, \\
    & V^3 (s)= v_3 \cos \left(\frac{s (p-2 q) (v_1-v_2)}{p}\right)+v_4 \sin \left(\frac{s (p-2 q) (v_1-v_2)}{p}\right), \\
    & V^4(s) = v_4 \cos \left(\frac{s (p-2 q) (v_1-v_2)}{p}\right)-v_3 \sin \left(\frac{s (p-2 q) (v_1-v_2)}{p}\right).
\end{align}

Equal penalty factors clearly correspond to the limit $p=q=1$.
In the more general setting of $q\not=p$, the length of the 
geodesics characterized by the $v_i$ is given by:
\begin{align}
    \int_0^1 \sqrt{G_{IJ}V^I V^J}ds = \sqrt{q(v_1^2+v_2^2)+p(v_3^2+v_4^2)}.
\end{align}

Substituting the obtained $V^I(s)$ and keeping only the leading order term 
in the Dyson series, $U(s)$ can be written as: 
\begin{align}
    U(s) \approx \exp(-i(\beta_1(s)M_1+ \beta_2(s)M_2 + \beta_3(s) M_3+ \beta_4(s) M_4)),
\end{align}
where the $\beta_I$ are as follows
\begin{align}
    \beta_1 &= s v_1, \\
    \beta_2 &= s v_2, \\ 
    \beta_3 &= \frac{1}{(p-2q)(v_1-v_2)}\bigg( -p v_4 + p v_4 \cos \left(\frac{s (p-2 q) (v_1-v_2)}{p}\right) \nonumber \\ 
    &+ p v_3 \sin \left(\frac{s (p-2 q) (v_1-v_2)}{p}\right) \bigg) \\ 
    \beta_4 &= \frac{1}{(p-2q)(v_1-v_2)}\bigg( p v_3 - p v_3 \cos \left(\frac{s (p-2 q) (v_1-v_2)}{p}\right) \nonumber\\ 
    &+ p v_4 \sin \left(\frac{s (p-2 q) (v_1-v_2)}{p}\right) \bigg).
\end{align}

The final condition $U(s=1) = U_{\rm target}$ implies:
\begin{align}
    \exp(-i(\beta_1(1)M_1+ \beta_2(1)M_2 + \beta_3(1) M_3+ \beta_4(1) M_4))= \exp(-i (\omega_1 M_1+\omega_2 M_2 + \mu^2 M_3)t),
\end{align}
such that:
\begin{align}
    \beta_1(1)= \omega_1 t, ~~~~ \beta_2(1)= \omega_2 t, ~~~~ \beta_3(1)= \mu^2 t, ~~~~ \beta_4(1)= 0. 
\end{align}

These equations are solved by:
\begin{align}
    v_1&=\omega_1 t,\\
    v_2&=\omega_2 t,\\
    v_3&=\mu^2t \frac{p-2q}{2p}(v_1-v_2)\cot\left(\frac{p-2q}{2p}(v_1-v_2)\right),\\
    v_4&=\mu^2t\frac{p-2q}{2p}(v_1-v_2)\,.
\end{align}

Using 
\begin{equation}
    v_1M_1+v_2M_2=\frac{1}{2}\left((v_1+v_2)(M_1+M_2)+ (v_1-v_2)(M_1-M_2)\right),
\end{equation}
and the $2\pi$-periodicity in the exponentiated $M_1+M_2$ and $M_1-M_2$, we need 
a period of $4\pi$ for $v_1-v_2$ and $v_1+v_2$. Therefore: 
\begin{equation}
    v_1\pm v_2=|(\omega_1 \pm \omega_2)t-4\pi\lfloor((\omega_1\pm \omega_2)t+2\pi)/(4\pi)\rfloor|.
\end{equation}

Finally, the complexity bound is given by:
\begin{align}
    &C_{[U_{\rm target}]}(\omega_1,\omega_2,\mu,t)\leq \sqrt{v_1^2+v_2^2+v_3^2+v_4^2}\\
    &= \sqrt{\frac{1}{2}\left((v_1+v_2)^2+(v_1-v_2)^2\right)+ \mu^4t^2\frac{(p-2q)^2}{4p^2}\frac{(v_1-v_2)^2}{\sin^2((p-2q)(v_1-v_2)/(2p))}}\,. \nonumber
\end{align}
Examples are shown in Figs.~\ref{fig3a} and \ref{fig3b}.
For $\omega_1=\omega_2$, the result simplifies to
\begin{equation}
    C_{\rm U_{\rm target}]}(\omega,\omega,\mu,t) 
    \leq \sqrt{\frac{1}{2}(v_1+v_2)^2+\mu^4t^2},
\end{equation}
and is independent of $q$ and $p$. Here,
\begin{equation}
    v_1+v_2=2|\omega t-2\pi\lfloor(\omega t+\pi)/2\pi\rfloor|\,.
\end{equation}

\begin{figure}[h!]
    \centering
    \includegraphics[scale=0.6]{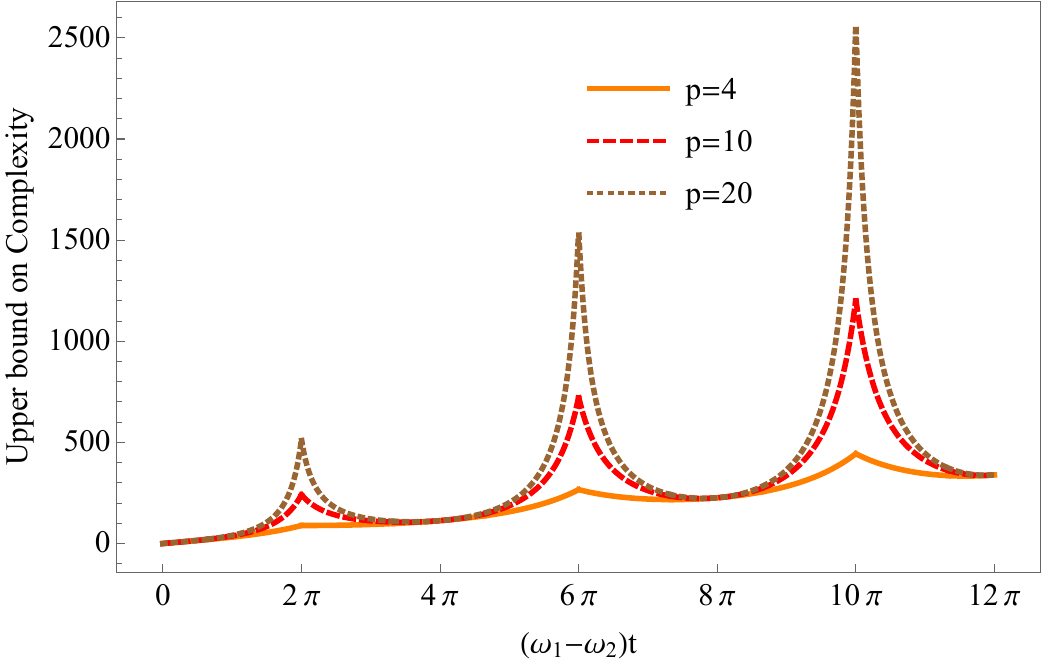}
    \caption{Behavior of complexity bounds for the time evolution operator of two coupled oscillators as a function of time for different values of the penalty factor $p$. The frequencies of the two oscillators are fixed at $\omega_1=2$ and $\omega_2=1$, and the coupling constant is fixed at $\mu=3$. The penalty factor $q$ is fixed at 1.}
    \label{fig3a}
\end{figure}

\begin{figure}[h!]
    \centering
    \includegraphics[scale=0.55]{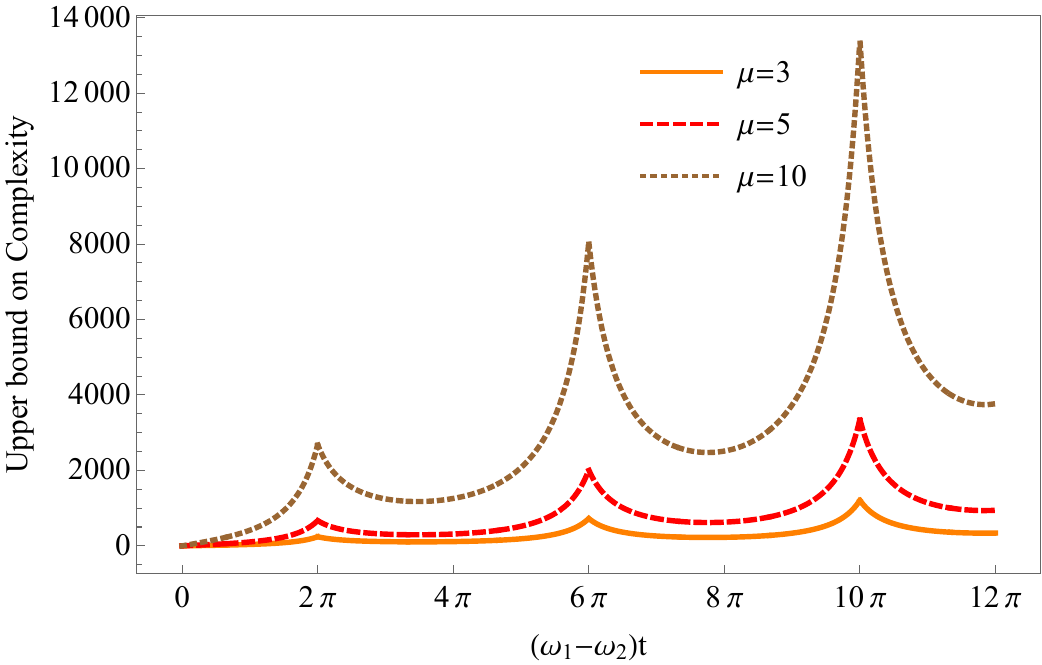}
    \caption{Behavior of complexity bounds for the time evolution operator of two coupled oscillators as a function of time for different values of the coupling constant $\mu$. The frequencies of the two oscillators are fixed at $\omega_1=2$ and $\omega_2=1$. The plots have been made by choosing $p=10$ and $q=1$.}
    \label{fig3b}
\end{figure}

\section{Anharmonic oscillator with the cubic term}

With an additional ingredient in the motivation of penalty factors, 
our methods can be applied even to anharmonic systems. As an example, 
we choose the Hamiltonian:
\begin{align}
    \textbf{H}_{A.HO}= \omega\bigg(\frac{P^2}{2}+ \frac{Q^2}{2}\bigg)+ \lambda Q^3.
    \label{QubicHamiltonian}
\end{align}
In this case, the generators can be considered to be the original: 
\begin{equation}
    M_1=\frac{1}{2}Q^2+\frac{1}{2}P^2,\quad M_2=\frac{1}{2}Q^2-\frac{1}{2}P^2,
    \quad M_3=\frac{1}{2}(QP+PQ),
\end{equation}
already used for the harmonic oscillator, together with: 
\begin{equation}
    M_4=Q^3,\quad M_5=P^3,\quad M_6=Q^2P+QPQ+PQ^2,
    \quad M_7=QP^2+PQP+P^2Q,
    \end{equation}
and so on with higher powers in $Q$ and $P$.

In terms of these generators, the Hamiltonian (\ref{QubicHamiltonian}) 
can be written as: 
\begin{align}
    \textbf{H}_{A.HO}= \omega M_1 + \lambda M_4,
\end{align}
defining the target unitary:
\begin{align}
\label{targetanharmoniccubic}
    U_{\rm target}= \exp(-i (\omega M_1+\lambda M_4)t).
\end{align}

The generators $M_2$ and $M_3$ do not appear in commutators of 
the other ones, $M_1$, $M_4$, $M_5$, $M_6$, $M_7$. They are, 
therefore, not required for Euler--Arnold equations with the 
desired target unitary. Commutators of the latter generators 
are given by:
\begin{align}
    [M_1,M_4] &= -i(PQ^2+QPQ+Q^2P)=-iM_6,\\
     [M_1,M_6] &= -2i(P^2Q+PQP+QP^2)+3iQ^3= -2i M_7 +3i M_4,\\
    [M_1,M_7] &= 2i(Q^2P+ QPQ+PQ^2)-3iP^3= 2iM_6 -3i M_5, \\
    [M_1,M_5] &= i(QP^2+PQP+PQ^2)=i M_7,\\
    [M_4,M_5] &= 3i (Q^2P^2+PQ^2P+P^2Q^2),\\
    [M_4,M_6] &= 9i Q^4,\\
    [M_4,M_7] &= 6i Q^3P+6iPQ^3+3i QPQ^2+3i Q^2PQ,\\
    [M_5,M_6] &= \mathcal{O}(M_I^4),\\
    [M_5,M_7] &= \mathcal{O}(M_I^4),\\
    [M_6,M_7] &= \mathcal{O}(M_I^4),
\end{align}
where the higher-order terms in the last three equations 
require the introduction of additional independent generators. 
Using those new generators in commutators with the original 
ones requires even higher orders. Iterating this procedure 
does not result in a finite-dimensional closed algebra 
suitable for this system.

Using suitable penalty factors, we can nevertheless propose 
a method to deal with target unitary operators whose generators 
are not part of a finite-dimensional closed commutator algebra. 
Illustrating the general method for the example just introduced, 
our proposal consists of two steps: 
\begin{itemize}
    \item \textit{Step 1}: The generators of order higher than that appearing in the target unitary operators are assigned prohibitively large penalties, such that geodesics will not move in their direction on the operator space.
    We can neglect their contribution to the Euler--Arnold equation and the resulting complexity.
    
    For example, in the commutator algebra shown above, we will assign 
    prohibitive penalties to the generators with quartic or higher powers, 
    such as $Q^4$, $Q^2P^2$, and $PQ^3$. For purposes of geodesic distance, 
    their contributions to the algebra are therefore neglected. 
    Under this assumption, the Euler--Arnold equations can be written as: 
    \begin{align}
        \frac{dV^1}{ds} &\approx -\frac{G_{66}}{G_{11}}V^4 V^6- 2\frac{G_{77}}{G_{11}} V^6V^7+ 3\frac{G_{44}}{G_{11}}V^4 V^6 \nonumber \\
        &+2\frac{G_{66}}{G_{11}}V^7 V^6-3 \frac{G_{55}}{G_{11}}V^7 V^5+ \frac{G_{77}}{G_{11}}V^5 V^7, \\
        \frac{dV^4}{ds} &\approx  \frac{G_{66}}{G_{44}}V^1V^6, \\
        \frac{dV^5}{ds} &\approx -\frac{G_{77}}{G_{55}}V^1V^7,\\
        \frac{dV^6}{ds} &\approx 2\frac{G_{77}}{G_{66}}V^1V^7-3\frac{G_{44}}{G_{66}}V^1V^4, \\
        \frac{dV^7}{ds} &\approx -2\frac{G_{66}}{G_{77}}V^1V^6+3\frac{G_{55}}{G_{77}}V^1V^5.
    \end{align}
    \item \textit{Step 2}: The generators of higher order have comparatively much higher penalties, i.e., the generators $M_4$, $M_5$, $M_6$, and $M_7$ which 
    are of cubic order have sufficiently large penalties compared to $M_1$, which 
    is quadratic.
\end{itemize}

A large penalty factor, such as $G_{66}$, means that a geodesic will not 
move in the corresponding direction, and therefore the component $V^6$ 
remains small. We can then ignore products of the prohibited components 
even if they are multiplied by a large penalty factor, such as $(G_{66}/G_{11})V^6V^7$, as well as terms with a single factor of a 
prohibited component as long as it is not multiplied by a large penalty 
factor, such as $(G_{55}/G_{11})V^5V^7$. Following this procedure, the 
equations simplify slightly to:
\begin{align}
        \frac{dV^1}{ds} &\approx -\frac{G_{66}}{G_{11}}V^4 V^6 + \frac{G_{77}}{G_{11}}V^5 V^7, \\
        \frac{dV^4}{ds} &\approx  \frac{G_{66}}{G_{44}}V^1V^6, \\
        \frac{dV^5}{ds} &\approx -\frac{G_{77}}{G_{55}}V^1V^7,\\
        \frac{dV^6}{ds} &\approx 2\frac{G_{77}}{G_{66}}V^1V^7-3\frac{G_{44}}{G_{66}}V^1V^4, \\
        \frac{dV^7}{ds} &\approx -2\frac{G_{66}}{G_{77}}V^1V^6+3\frac{G_{55}}{G_{77}}V^1V^5,
    \end{align} 
    if $M_6$ and $M_7$ have prohibitive penalties. If $M_5$ has a prohibitive penalty as well, we obtain
    \begin{align}
        \frac{dV^1}{ds} &\approx -\frac{G_{66}}{G_{11}}V^4 V^6, \\
        \frac{dV^4}{ds} &\approx  \frac{G_{66}}{G_{44}}V^1V^6, \\
        \frac{d(G_{55}V^5)}{ds} &\approx -G_{77}V^1V^7,\\
        \frac{d(G_{66}V^6)}{ds} &\approx 2G_{77}V^1V^7-3G_{44}V^1V^4, \\
        \frac{d(G_{77}V^7)}{ds} &\approx -2G_{66}V^1V^6+3G_{55}V^1V^5.
    \end{align} 

Similarly, if $M_4$ has prohibitive penalties, we can ignore the 
term $V^4V^6$ and simplify the Euler--Arnold equations further.
We will consider the penalties $G_{44}=G_{55}=G_{66}=G_{77}=p \gg G_{11}$. 
In that case, the Euler--Arnold equations simplify to: 
\begin{align}
    \frac{dV^1}{ds} &\approx 0, \\
    \frac{dV^4}{ds} &\approx V^1V^6, \\
    \frac{dV^5}{ds} &\approx -V^1V^7,\\
    \frac{dV^6}{ds} &\approx 2V^1V^7- 3 V^1V^4, \\
    \frac{dV^7}{ds} &\approx -2V^1V^6+ 3 V^1V^5.
\end{align}

The solutions of the equations can be written as:
\begin{align}
    V^1(s) &= v_1,  \\ \nonumber
    V^4(s) &= \frac{1}{4} v_4 (3 \cos (s v_1)+\cos (3 s v_1))+\frac{1}{4} v_5 (3 \sin (s v_1)-\sin (3 s v_1)) \\ &+\frac{1}{4} v_6 (\sin (s v_1)+\sin (3 s v_1))+\frac{1}{4} v_7 (\cos (s v_1)-\cos (3 s v_1)),  \\ \nonumber
    V^5(s) &= \frac{1}{4} v_4 (\sin (3 s v_1)-3 \sin (s v_1))+\frac{1}{4} v_5 (3 \cos (s v_1)+\cos (3 s v_1))~\\ &+\frac{1}{4} v_6 (\cos (s v_1)-\cos (3 s v_1))+\frac{1}{4} v_7 (-\sin (s v_1)-\sin (3 s v_1)), \\ \nonumber
    V^6(s) &= -\frac{3}{4} v_4 (\sin (s v_1)+\sin (3 s v_1))+\frac{3}{4} v_5 (\cos (s v_1)-\cos (3 s v_1)) \\ &+\frac{1}{4} v_6 (\cos (s v_1)+3 \cos (3 s v_1))+\frac{1}{4} v_7 (3 \sin (3 s v_1)-\sin (s v_1)), \\ \nonumber
    V^7(s) &= \frac{3}{4} v_4 (\cos (s v_1)-\cos (3 s v_1))+\frac{3}{4} v_5 (\sin (s v_1)+\sin (3 s v_1)) \\ &+ \frac{1}{4} v_6 (\sin (s v_1)-3 \sin (3 s v_1))+\frac{1}{4} v_7 (\cos (s v_1)+3 \cos (3 s v_1)).
\end{align}

Proceeding as before, we would keep only up to the leading order term in the 
Dyson series, which helps us to write the path-ordered exponential as follows:
\begin{align}
    U(s) \approx \exp(-i(\gamma_1(s) M_1+ \gamma_4(s) M_4+ \gamma_5(s) M_5 + \gamma_6(s) M_6 + \gamma_7(s) M_7)),
\end{align}
where the following formulas give the $\gamma_I$ functions:
\begin{align}
    \gamma_1(s) & = s v_1, \\ \nonumber
    \gamma_4(s) &= \frac{1}{12 v_1}\bigg(9 v_4 \sin (s v_1)+v_4 \sin (3 s v_1)-3 (3 v_5+v_6) \cos (s v_1)+(v_5-v_6) \cos (3 s v_1) \\ &+3 v_7 \sin (s v_1)-v_7 \sin (3 s v_1)+8 v_5+4 v_6\bigg), \\ \nonumber
    \gamma_5(s) &= \frac{1}{12 v_1} \bigg(3 (3 v_4+v_7) \cos (s v_1)+(v_7-v_4) \cos (3 s v_1)+9 v_5 \sin (s v_1) \\ & +v_5 \sin (3 s v_1)+3 v_6 \sin (s v_1)-v_6 \sin (3 s v_1)-8 v_4-4 v_7\bigg), \\ \nonumber
    \gamma_6(s) &= \frac{1}{4 v_1}\bigg((3 v_4+v_7) \cos (s v_1)+(v_4-v_7) \cos (3 s v_1)+3 v_5 \sin (s v_1)-v_5 \sin (3 s v_1) \\ &+v_6 \sin (s v_1)+v_6 \sin (3 s v_1)-4 v_4\bigg), \\ \nonumber
    \gamma_7(s) &= \frac{1}{4 v_1}\bigg(3 v_4 \sin (s v_1)-v_4 \sin (3 s v_1)-(3 v_5+v_6) \cos (s v_1)+(v_6-v_5) \cos (3 s v_1) \\ &+v_7 \sin (s v_1)+v_7 \sin (3 s v_1)+4 v_5\bigg).
\end{align}
}
Upon implementing the boundary condition at $s=1$ by setting $U(s=1)$ equal 
to the target unitary operator written in (\ref{targetanharmoniccubic}), 
we obtain:
\begin{align}
    \gamma_1(1)= \omega t, ~~~ \gamma_4(1)= \lambda t, ~~~ \gamma_5(1)=0, ~~~ \gamma_6(1)=0, ~~ \gamma_7(1)=0.
\end{align}

Our treatment of higher-order contributions by suppressing them through 
large penalty factors also simplifies the periodicity argument. The 
commutators of $M_1$ through $M_7$ with neglected fourth-order terms 
take the form of a semidirect product of the Lie algebra 
${\rm sp}(2,{\mathbb R})$ spanned by $M_1$, $M_2$ and $M_3$ with an 
Abelian 4-dimensional Lie algebra spanned by $M_4$ through $M_7$. 
The precise form of the semi-direct product can be determined by the 
same methods used in \cite{Bosonize} for the algebra formed by Poisson 
brackets of third-order central moments: It is given by ${\rm sp}(2,{\mathbb R})\ltimes{\mathbb R}^4$ where ${\rm sp}(2,{\mathbb R})$ acts on 
${\mathbb R}^4$ according to the spin-$3/2$ representation of 
${\rm sp}(2,{\mathbb R})$. Imposing periodicity can then be done 
by the same arguments used for a harmonic oscillator, based on 
properties of ${\rm sp}(2,{\mathbb R})$, requiring $4\pi$-periodicity 
of $v_1$ in $\omega t$:
\begin{equation}
    v_1=|\omega t-4\pi \lfloor(\omega t+2\pi)/(4\pi)\rfloor|\,.
\end{equation}
With this result, the other coefficients are:
\begin{eqnarray}
    v_4&=&\frac{3 \lambda  v_1 t \cos (v_1) \cot(v_1/2)}{2(1+2 \cos (v_1))},\\
    v_5&=& 0, \\
    v_6&=& \frac{3v_1 \lambda t}{2},\\
    v_7&=& \frac{3v_1 \lambda t \sin(v_1)}{2(1+ 2 \cos(v_1))}.
\end{eqnarray}

The initial velocity components $v_4$ and $v_7$ diverge at 
$\omega t= \pm \frac{2 n \pi}{3}$, which means that the complexity 
bound also diverges at these points. Since we are using a compact 
group for this system, it is exponential, and we are not missing 
parts of the manifold. The right-invariant metric used here does 
not seem to be complete. Around a divergence, neglecting higher-order 
terms in the full algebra may not be justified because, compared 
with an infinite distance, they could certainly contribute to 
geodesics even if they are subject to prohibitive penalties. If 
one removes suitable regions around the divergences, the finite 
geodesic distances are reliable as upper bounds on the complexity.

The complexity bound of the time evolution operator keeping up to 
cubic order terms can therefore be written as: 
\begin{align}
\nonumber
    C[U_{A.HO}] & \leq  \int_0^1 ds \sqrt{G_{IJ}V^I(s)V^J(s)} \\ \nonumber
     &= \int_0^1 ds \sqrt{G_{11}(V^1)^2+p((V^4)^2+(V^5)^2+(V^6)^2+(V^7)^2)}\\ \nonumber
    &= \int_{0}^1 ds \bigg(\frac{1}{2}\bigg[4 G_{11} v_1^2+p \cos (4 s v_1) \left(-3 v_4^2+2 v_4 v_7-3 v_5^2+2 v_5 v_6+v_6^2+v_7^2\right) \\  
    &+4 p \sin (4 s v_1) (v_5 v_7-v_4 v_6)+p \left(7 v_4^2-2 v_4 v_7+7 v_5^2-2 v_5 v_6+3 \left(v_6^2+v_7^2\right)\right)\bigg]^{1/2}\bigg).
\end{align}

The integrand is no longer constant, but it can be integrated upon using $A+B\cos(x)+C\sin(x)=A+\sqrt{B^2+C^2}\sin(x+\phi)$ with $\sin\phi=B/\sqrt{B^2+C^2}$ and $\int\sqrt{a+b\sin y}dy=-2\sqrt{a+b}E(\frac{1}{4}(\pi-2x)|2b/(a+b))$ with an elliptic function of the second kind. 
Here, we have:
\begin{align}
    A &= 2 G_{11}v_1^2 + \frac{p}{2}(7 v_4^2-2 v_4 v_7+7 v_5^2-2 v_5 v_6+3 (v_6^2+v_7^2)), \\
    B &= \frac{p}{2}(-3 v_4^2+2 v_4 v_7-3 v_5^2+2 v_5 v_6+v_6^2+v_7^2),\\
    C &= 2 p (v_5v_7-v_4v_6),\\
    x &= 4 s v_1,
\end{align}
with the above identification, $a$ and $b$ can be written as: 
\begin{align}
    a &= A= 2 G_{11}v_1^2 + \frac{p}{2}(7 v_4^2-2 v_4 v_7+7 v_5^2-2 v_5 v_6+3 (v_6^2+v_7^2)) \\
    b &= \sqrt{B^2+C^2}= \frac{p}{2}\sqrt{\left(-3 v_4^2+2 v_4 v_7-3 v_5^2+2 v_5 v_6+v_6^2+v_7^2\right)^2+16 (v_4 v_6-v_5 v_7)^2}\,.
\end{align}

\begin{figure}[h!]
    \centering
    \includegraphics[scale=0.6]{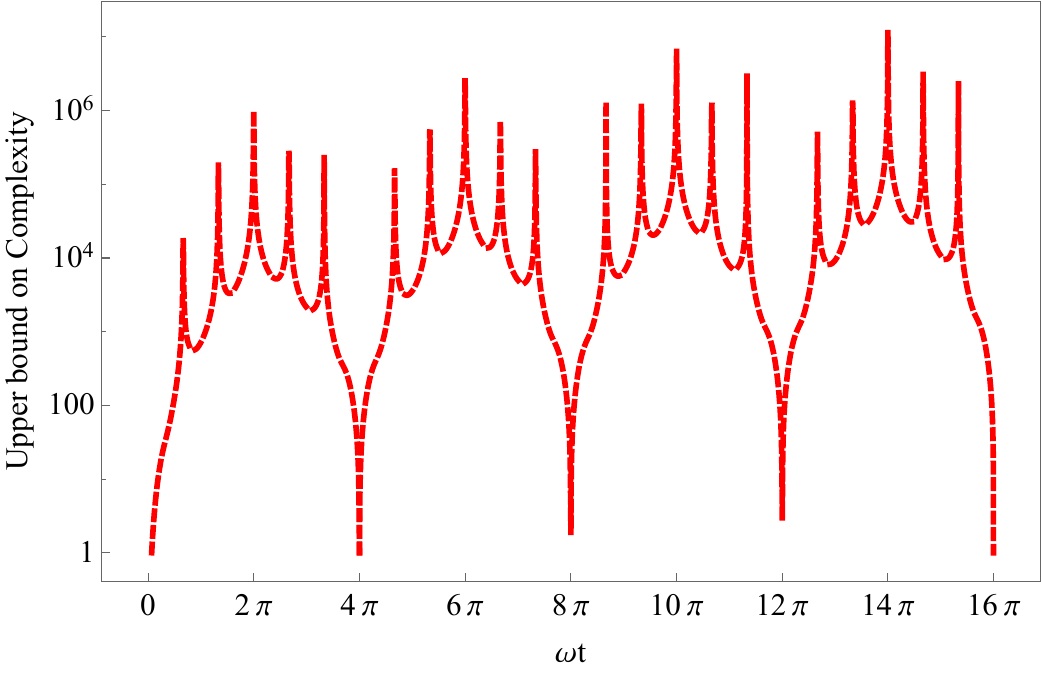}
    \caption{Complexity of the time evolution operator of an anharmonic oscillator. \label{figAn}}
\end{figure}

Substituting $y= x+\phi$, the complexity bound can be explicitly evaluated, as used in the plot for Fig.~\ref{figAn}. 

In the harmonic limit $\lambda\to0$ (with $G_{11}=1$ as used earlier), we have:
\begin{align}
    C[U_{A.HO}]\bigg|_{\lambda \rightarrow 0}= |\omega t-4\pi \lfloor(\omega t+2\pi)/(4\pi)\rfloor|\
\end{align}
in agreement with our direct derivation for the harmonic oscillator.
In this limit, the initial velocities $v_4$, $v_6$, and $v_7$ go to zero, 
and the only contribution comes from $v_1$.  

\section{Discussion and Comments}

The geometrical approach to quantum complexity is intriguing because 
it expresses a complicated optimization problem with a standard procedure 
of finding the geodesic distance between two given points in a suitable 
curved space, as proposed in \cite{Nielsen_2006,https://doi.org/10.48550/arxiv.quant-ph/0502070,https://doi.org/10.48550/arxiv.quant-ph/0701004}. In the application 
of geodesic distance to complexity, the two endpoints are represented by the
identity and a target unitary operator, respectively, embedded in a suitable
geometrical formulation of the group of unitary operators on a Hilbert space. 
On a given Lie group equipped with a right-invariant metric, the problem of 
finding geodesics, whose lengths then determine the complexity of operations, 
is reduced to solving the Euler--Arnold equations. 

Initial subtleties that immediately arise in an infinite-dimensional 
setting can be evaded by using a truncation of the full group of unitary 
operators to a suitable finite-dimensional subgroup that includes operators 
of interest, including the target unitary as well as additional basic 
operations that form a closed group together with the target unitary.
In this spirit, we have revisited the question of the quantum complexity 
of the harmonic oscillator within various finite-dimensional groups, 
which had already been discussed in other papers, in order to understand 
its properties and with an aim toward possible generalizations to related
evolution operators. In this process, we found and highlighted several 
additional mathematical subtleties, even in a finite-dimensional construction 
that require a proper consideration of interesting properties of the group 
theory. In particular, the question of properly embedding the 
finite-dimensional truncation in the full group of unitary operators, 
which had not been considered before, requires information about topological 
properties of Lie groups and their covering groups.

We studied the harmonic oscillator from the perspective of two different Lie
groups, the solvable harmonic oscillator group and the semisimple 
${\rm Sp}(2,{\mathbb R})$. Our results qualitatively agree with previous 
work, showing a piecewise linear oscillating behavior of the quantum complexity 
of the evolution operator. However, based on the condition that the finite-
dimensional group should be embeddable in the full infinite-dimensional group 
of unitary operators with the well-known spectrum for the evolution operator 
of the harmonic oscillator, we concluded that the period of oscillations should 
be doubled. This condition takes into account the half-integer nature of the
ground-state energy. The correct group is, therefore not directly 
${\rm Sp}(2,{\mathbb R})$ but rather its 2-fold covering, the metaplectic 
group ${\rm Mp}(2,{\mathbb R})$. This group is not a matrix group, and 
therefore previous methods which explicitly used finite-dimensional 
matrix representations cannot be applied. Our methods, by contrast, 
work mainly at an algebraic level and bring in group-theoretical 
properties only in the final step in order to determine the periodicity 
properties. Our new methods, therefore, present crucial generalizations 
and clarifications even for the well-studied harmonic oscillator. 
While doubling the period does not change the short-term behavior of 
the complexity while the evolution operator is still close to the identity, 
it can have large effects after several periods of the system.

We modified the original Hamiltonian by adding terms linear or quadratic 
in the position operator, which can be done without enlarging the original 
finite-dimensional groups. Classically, such a procedure merely shifts 
the origin of the oscillator or changes the frequency, but there are 
stronger effects in quantum mechanics. In particular, rewriting the
added operator as a shift in one of the real-valued parameters might require
operations that are not contained in the finite-dimensional group chosen 
for calculations, or the addition might implicitly change penalty factors 
assigned to operations by choosing a specific right-invariant metric on 
the group. As we found, the quantum complexity may well depend on such 
innocuous-looking modifications. In particular, such terms might change 
the piecewise linear nature of the complexity of a harmonic oscillator.
The inverted harmonic oscillator can be treated with similar methods. 
Since its Hamiltonian corresponds to a non-compact direction in the 
finite-dimensional group used here, the complexity is not periodic 
and increases linearly, as seen in previous work.

We generalized the methods in two ways: by using larger but still 
finite-dimensional groups and estimating the complexity of two coupled 
oscillators and of an anharmonic oscillator. We were able to formulate 
these problems within subgroups with easily identifiable periodicity 
properties. In all cases, we encountered additional subtleties because 
the groups involved are not exponential, and because the right-invariant metrics 
motivated by physical arguments are not guaranteed to be geodesically 
complete. These properties, together with the overall truncation to 
a finite-dimensional group, imply that a given calculation does not 
take into account directions in the full infinite-dimensional group 
of unitary operators through which a geodesic might be able to take 
a shortcut. Such results can therefore be considered only upper 
bounds on the quantum complexity rather than strict values. This 
caveat applies in particular to divergences of length that 
could be implied by geodesic incompleteness. 

We encountered non-exponential groups and possible geodesic 
incompleteness because useful groups for bosonic oscillator systems 
are not compact. This property is different for fermionic oscillators, 
which lead to orthogonal groups generated by fermion bilinears. 
Some of the derivations may, therefore simplify for fermions, 
which we are planning to analyze in future work.

\acknowledgments

SC would like to thank the doctoral school of Jagiellonian University 
for providing a fellowship during the course of the work. The research 
was conducted within the Quantum Cosmos Lab (\href{https://quantumcosmos.org}
{https://quantumcosmos.org}) at the Jagiellonian University. We would 
like to thank Mario Flory, Bret Underwood, Shajid Haque, Ghadir Jafari, \v{S}\'arka Blahnik,  Suddhasattwa Brahma, Sebastien MacDonald and Daniel Quiles Pastor
for their insightful comments on our draft. MB was supported in part by 
NSF grant PHY-2206591.

\begin{appendix}
\section{Line element}
\label{appA}

In order to derive the expression for a right-invariant line element 
on a Lie group with generators $\mathcal{O}_I$, we first compute the 
differential:
\begin{equation} \label{dU}
    dU = -i \mathcal{O}_I U dx^I,
\end{equation}
of a group element $U = \exp(-i x^I \mathcal{O}_I)$ that exponentiates 
a fixed generator $\mathcal{O}_I$ (no summation over $I$). The line 
element requires us to solve for $dx^I$, for which we first multiply 
(\ref{dU}) with $U^{-1}$ from the right:
\begin{equation}
    i dU U^{-1} = \mathcal{O}_I dx^I\,.
\end{equation}

The generator $\mathcal{O}_I$ does not have an inverse, but if we 
use a matric representation of the Lie algebra, we can first multiply 
by $\mathcal{O}_I^{\dagger}$, take a trace in the representation, 
and divide the equation by the resulting number:
\begin{equation} \label{dx}
    dx^I = \frac{1}{{\rm Tr}[\mathcal{O}_I \mathcal{O}_I^{\dagger}]}\bigg[{\rm Tr}[i dU U^{-1} \mathcal{O}_I^{\dagger}]\bigg]\,.
\end{equation}
(If the trace happens to be zero, it effectively puts a prohibitive 
penalty on the corresponding generator.)

The generator $\mathcal{O}_I$ commutes with $U$, and any ordering 
could have been chosen on the right of (\ref{dU}). The way in which 
we wrote (\ref{dU}) is suitable for the construction of a right-invariant 
line element by transporting the resulting $dx^I$ to generic group 
elements: The product $dU U^{-1}$ is right-invariant because for any 
constant $g$ in the Lie group, $d(Ug) (gU)^{-1}=dU U^{-1}$. After such 
a right translation on the group, the ordering is relevant because 
$d(Ug)$ is no longer guaranteed to commute with $(Ug)^{-1}$, unlike
$dU$ and $U^{-1}$ for our specific $U$.

After the right transportation, we can use (\ref{dx}) at any group 
element $U$. Using the penalty matrix $G_{IJ}$, the line element 
on the entire group then takes the form: 
\begin{align}
    ds^2 &= G_{IJ}dx^{I}dx^{J}= G_{IJ}\frac{1}{{\rm Tr}[\mathcal{O}_I \mathcal{O}_I^{\dagger}]}\bigg[{\rm Tr}[i dU U^{-1} \mathcal{O}_I^{\dagger}]\bigg]\frac{1}{{\rm Tr}[\mathcal{O}_J \mathcal{O}_J^{\dagger}]}\bigg[{\rm Tr}[i dU U^{-1} \mathcal{O}_J^{\dagger}]\bigg] \nonumber \\
    &= G_{IJ}\frac{1}{{\rm Tr}[\mathcal{O}_I \mathcal{O}_I^{\dagger}]}\frac{1}{{\rm Tr}[\mathcal{O}_J \mathcal{O}_J^{\dagger}]}\bigg[{\rm Tr}[i U^{-1} \mathcal{O}_I^{\dagger} dU]\bigg]\bigg[{\rm Tr}[i U^{-1} \mathcal{O}_J^{\dagger}dU]\bigg],  
\end{align}
where we brought $dU$ to the right using cyclic commutation in the trace. 

\section{Using a generic element of the suitable group to compute complexity}
\label{appB}

In this appendix, we show that instead of trying to write down the 
Dyson series for the path-ordered exponential, we can make use of 
the differential equation (\ref{differentialU}) it satisfies in 
order to derive the complexity. In fact, this method is usually 
adopted in the circuit complexity literature. Let us explain the 
steps usually followed, taking the example of the model of two 
coupled oscillators studied in \cite{Jefferson:2017sdb}. It was 
realized that for the given purpose, it is necessary to look for 
geodesics in the ${\rm GL}(2,\mathbb{R})$ group. One may choose an explicit 
parametrization of a general element $U \in {\rm GL}(2,\mathbb{R})$, equation 
(3.18) of \cite{Jefferson:2017sdb}, and a suitable finite-dimensional 
matrix representation of the generators. However, in general, 
there may be obstacles to finding a finite-dimensional matrix 
representation of the generators, for instance, for certain 
universal covering groups. In this appendix, we, therefore, 
use an algebraic method based on a suitable parametrization 
of a general element of the desired group without using any
matrix representation of the generators. 
 
 We will explicitly calculate the complexity of the 
 displacement operator using the product form (disentangled form) of 
 the generic element of the Harmonic oscillator group. To remind, our 
 intention is to solve: 
 \begin{align}
 \label{duapp}
     \frac{dU(s)}{ds}= -i V^I(s) \mathcal{O}_I U(s),
 \end{align}
subject to the boundary conditions: 
\begin{align}
    U(s=0)= \mathbb{I}, ~~~~ {\rm and}~~~ U(s=1)= U_{\rm target}.
\end{align}

To solve the above equation, we discussed that we need to introduce 
a generic element of the Lie group under consideration, and there
are two ways to represent the generic element:
\begin{align}
    U(s)= \exp\left(-i \sum_{i=1}^n \beta_i(s) \mathcal{\hat{O}}_i \right)= \prod_{i=1}^{N} \exp(-i \alpha_i(s) \mathcal{\hat{O}}_i).
\end{align}

In the main text, we have worked with the form 
$\exp(-i \sum_{i=1}^n \beta_i(s) \mathcal{\hat{O}}_i$). 
Here we do the computation for the product form, i.e. 
we take:
\begin{align}
    U(s)= \prod_{i=1}^{N} \exp(-i \alpha_i(s) \mathcal{\hat{O}}_i).
\end{align}

For the Harmonic oscillator group with generators $H$, $P$, $Q$ and $E$, the generic element can be written as: 
\begin{align}
\label{genericHOapp}
    U(s)= \exp(-i \alpha_1(s) E)\exp(-i \alpha_2(s) P)\exp(-i \alpha_3(s)Q)\exp(-i \alpha_4 H).
\end{align}

Substituting the above equation in \ref{duapp}, the LHS can be written as:
\begin{align}
\label{LHSapp}
\nonumber
    \frac{dU(s)}{ds} &= -i \biggl[\exp\{-i \alpha_1(s) E\}\exp\{-i \alpha_2(s) P\}\exp\{-i \alpha_3(s) Q\} \bigg( \exp\{-i \alpha_4(s)H\} \alpha_4'(s) H\bigg) \\ \nonumber &
    +\exp\{-i \alpha_1(s) E\}\exp\{-i \alpha_2(s) P\}\bigg( \exp\{-i \alpha_3(s) Q\}  \alpha_3'(s)Q\bigg)\exp\{-i \alpha_4(s) H\} \\ \nonumber& 
    + \exp\{-i \alpha_1(s) E\}\bigg( \exp\{-i \alpha_2(s) P\} \alpha_2'(s) P\bigg)\exp\{-i \alpha_3(s) Q\}\exp\{-i \alpha_4(s) H\} \\  & 
    + \bigg( \exp\{-i \alpha_1(s) E\} \alpha_1'(s) E \bigg)\exp\{-i \alpha_2(s) P\}\exp\{-i \alpha_3(s) Q\}\exp\{-i \alpha_4(s) H\}\biggl],
\end{align}
whereas the RHS (for $G_{IJ}=\delta_{IJ}$) can be written as: 
\begin{align}
\label{RHSapp}
\nonumber
    -i V^I(s) \mathcal{O}_I U(s)
    & = -i \bigg(v_H H + \bigg\{v_P \cos (s (v_E-v_H))+v_Q \sin (s (v_E-v_H))\bigg\} P \\ & ~~~~ + \bigg\{v_Q \cos (s (v_E-v_H))-v_P \sin (s(v_E- v_H))\bigg\} Q + v_E E\bigg) U(s).
\end{align}

Before proceeding, let us make a simplification by realizing the action 
of the generator $E$. The generator $E$ being a non-trivial center, commutes
with all other generators, and its role is just to produce a phase factor 
when applied to states. Therefore to simplify, we can choose $v_E=0$ for 
which eqn \ref{RHSapp} becomes: 
\begin{align}
    -i V^I(s) \mathcal{O}_I U(s)
    & = -i \bigg(v_H H + \bigg\{v_P \cos (s (v_H))-v_Q \sin (s (v_H))\bigg\} P \\ & ~~~~ + \bigg\{v_Q \cos (s (v_H))+v_P \sin (s( v_H))\bigg\} Q\bigg) U(s).
\end{align}

To equate the coefficients of the generators in \ref{LHSapp} and 
\ref{RHSapp} and derive the differential equations for the $s$ 
dependent parameters $\alpha_i$'s, we have to express \ref{LHSapp} 
in way such that it resembles \ref{RHSapp} in its form i.e it should 
be written as $(~~~~) \times U(s)$. For that purpose, let us begin by 
writing equation \ref{LHSapp} as: 
\begin{align}
\label{LHSterms}
    \frac{dU(s)}{ds} &= -i \bigg[{\rm Term1}+{\rm Term2}+{\rm Term 3}+{\rm Term 4}\bigg],
\end{align}
where {\rm Term 1}, {\rm Term 2}, {\rm Term 3}, and {\rm Term 4} are as follows:
\begin{align}
    {\rm Term 1} &= \exp\{-i \alpha_1(s) E\}\exp\{-i \alpha_2(s) P\}\exp\{-i \alpha_3(s) Q\} \bigg( \exp\{-i \alpha_4(s)H\} \alpha_4'(s) H\bigg), \\
    {\rm Term 2} &= \exp\{-i \alpha_1(s) E\}\exp\{-i \alpha_2(s) P\}\bigg( \exp\{-i \alpha_3(s) Q\}  \alpha_3'(s)Q\bigg)\exp\{-i \alpha_4(s) H\}, \\
    {\rm Term 3} &= \exp\{-i \alpha_1(s) E\}\bigg( \exp\{-i \alpha_2(s) P\} \alpha_2'(s) P\bigg)\exp\{-i \alpha_3(s) Q\}\exp\{-i \alpha_4(s) H\}, \\ 
    {\rm Term 4} &= \bigg( \exp\{-i \alpha_1(s) E\} \alpha_1'(s) E \bigg)\exp\{-i \alpha_2(s) P\}\exp\{-i \alpha_3(s) Q\}\exp\{-i \alpha_4(s) H\}.
\end{align}

Let us start the simplification with the easiest case, which is Term 4.
We can easily shift the generator $E$ to the left of $\exp\{-i \alpha_1(s) E\}$ 
as $E$ commutes with it. Thus, Term 4 can be written as:
\begin{align}
\nonumber
    {\rm term 4} &= \alpha_1'(s) E  \exp\{-i \alpha_1(s) E\}\exp\{-i \alpha_2(s) P\}\exp\{-i \alpha_3(s) Q\}\exp\{-i \alpha_4(s) H\} \\
    & = \alpha_1'(s) E U(s).
\end{align}
Now let us consider Term 3. Again, since $P$ commutes with 
$\exp\{-i \alpha_2(s)P\}$, we can shift $P$ to the left of 
the exponential i.e. we can rewrite Term 3 as:  
\begin{align}
\nonumber
    {\rm Term 3} &= \alpha_2'(s)\bigg(\exp\{-i \alpha_1(s) E\}P\bigg) \exp\{-i \alpha_2(s) P\}\exp\{-i \alpha_3(s) Q\}\exp\{-i \alpha_4(s) H\}  \\
    & = \alpha_2'(s)\bigg([\exp\{-i \alpha_1(s) E\},P]+P \exp\{-i \alpha_1(s) E\}\bigg) \exp\{-i \alpha_2(s) P\} \nonumber \\ & ~~~~~~~~~\exp\{-i \alpha_3(s) Q\}\exp\{-i \alpha_4(s) H\}.
\end{align}

So, we have to evaluate the commutator of $\exp\{-i \alpha_1(s) E\}$ with 
$P$. Since $E$ commutes with $P$, all the higher orders of $E$ will also 
commute with $P$, therefore: 
\begin{equation}
    [\exp\{-i\alpha_1(s)E\},P] = 0,
\end{equation}
and in consequence:
\begin{align}
\nonumber
    {\rm Term 3} &= \alpha_2'(s) P \exp\{-i \alpha_1(s)E\}\exp\{-i \alpha_2(s)P\}\exp\{-i \alpha_3(s)Q\}\exp\{-i \alpha_4(s)H\} \\
    & = \bigg(\alpha_2'(s) P\bigg) ~  U(s).
\end{align}

Now let us try to simply Term 2:
\begin{align}
\nonumber
    {\rm Term 2} &= \exp\{-i \alpha_1(s) E\}\exp\{-i \alpha_2(s) P\}\bigg( \exp\{-i \alpha_3(s) Q\}  \alpha_3'(s)Q\bigg)\exp\{-i \alpha_4(s) H\} \\ 
    & = \alpha_3'(s) \exp\{-i \alpha_1(s) E\}\bigg(\exp\{-i \alpha_2(s) P\}Q\bigg) \exp\{-i \alpha_3(s) Q\}\exp\{-i \alpha_4(s) H\}.
 \end{align}
 
We manipulate the term $\bigg(\exp\{-i \alpha_2(s) P\}Q\bigg)$ to shift 
the generator $Q$ to the left of the exponential. For this purpose, 
let us write $\bigg(\exp\{-i \alpha_2(s) P\}Q\bigg)$ as:
\begin{align}
\nonumber
    \bigg(\exp\{-i \alpha_2(s) P\}Q\bigg) &= \bigg(\exp\{-i \alpha_2(s) P\}Q\bigg) \underbrace{\exp\{i \alpha_2(s) P\}\exp\{-i \alpha_2(s) P\}}_{\text{Identity operator}} \\ 
    & = \bigg(\underbrace{\exp\{-i \alpha_2(s) P\}Q \exp\{i \alpha_2(s) P\}}_{\text{Apply Baker-Campbell Hausdorff}}\bigg)\exp\{-i \alpha_2(s) P\}.
\end{align}
Applying Baker-Campbell-Hausdroff (BCH) lemma:
\begin{align}
    e^{\lambda B} A e^{-\lambda B}= A+ \lambda [B,A]+ \frac{\lambda^2}{2!}[B,[B,A]]+...
\end{align}
 we can write the term $(\exp\{-i \alpha_2(s) P\}Q \exp\{i \alpha_2(s) P\})$ as: 
 \begin{align}
      \bigg(\exp\{-i \alpha_2(s) P\}Q \exp\{i \alpha_2(s) P\}\bigg) = Q- \alpha_2 E.
 \end{align}
 Therefore, Term 2 can be written as:
 \begin{align}
 \nonumber
   {\rm term 2} & = \alpha_3'(s) e^{-i \alpha_1(s) E} \bigg(Q- \alpha_2(s) E \bigg) e^{-i \alpha_2(s) P}e^{-i \alpha_3(s) Q}e^{-i \alpha_4(s) H} \\
    & = \alpha_3'(s) Q~ U(s) - \alpha_3'(s) \alpha_2(s) E~U(s).
\end{align}

Finally, let us simplify Term 1:
\begin{align}
\nonumber
     {\rm Term 1} &= \exp\{-i \alpha_1(s) E\}\exp\{-i \alpha_2(s) P\}\exp\{-i \alpha_3(s) Q\} \bigg( \exp\{-i \alpha_4(s)H\} \alpha_4'(s) H\bigg) \\ \nonumber
     & = \alpha_4'(s) \exp\{-i \alpha_1(s) E\}\exp\{-i \alpha_2(s) P\}\bigg(\exp\{-i \alpha_3(s) Q\} H\bigg)\exp\{-i \alpha_4(s)H\} \\ \nonumber
     & = \alpha_4'(s) e^{-i \alpha_1(s) E}e^{-i \alpha_2(s) P}\bigg(\exp\{-i \alpha_3(s) Q\} H\bigg)\underbrace{e^{i \alpha_3(s) Q}e^{-i \alpha_3(s) Q}}_{\text{Identity operator inserted}} \\ &~~~\exp\{-i \alpha_4(s)H\}\nonumber \\ \nonumber
     &= \alpha_4'(s) \exp\{-i \alpha_1(s) E\}\exp\{-i \alpha_2(s) P\}\bigg(\underbrace{\exp\{-i \alpha_3(s) Q\} H\exp\{i \alpha_3(s) Q\}}_{\text{Apply Baker-Campbell Hausdorff formula}}\bigg)\\ &~~~\exp\{-i \alpha_3(s) Q\} \exp\{-i \alpha_4(s)H\}\nonumber \\ \nonumber
     & =\alpha_4'(s) \exp\{-i \alpha_1(s) E\}\exp\{-i \alpha_2(s) P\}\bigg(H+\alpha_3(s) P+\frac{\alpha_3(s)^2}{2}E\bigg) \\ \nonumber & ~~~~~~~~\exp\{-i \alpha_3(s) Q\}\exp\{-i \alpha_4(s)H\}\\ \nonumber
    & = \alpha_4'(s) \exp\{-i \alpha_1(s) E\}\exp\{-i \alpha_2(s) P\} H \exp\{-i \alpha_3(s) Q\}\exp\{-i \alpha_4(s)H\} \\ \nonumber & ~~~
     + \alpha_4'(s) \exp\{-i \alpha_1(s) E\}\exp\{-i \alpha_2(s) P\} \alpha_3(s) P \exp\{-i \alpha_3(s) Q\}\exp\{-i \alpha_4(s)H\} \\ \nonumber & ~~~
     + \alpha_4'(s) \exp\{-i \alpha_1(s) E\}\exp\{-i \alpha_2(s) P\} \frac{\alpha_3(s)^2}{2}E \exp\{-i \alpha_3(s) Q\}\exp\{-i \alpha_4(s)H\} \\ \nonumber
     & = \alpha_4'(s) \exp\{-i \alpha_1(s) E\}\bigg(H-\alpha_2(s) Q+\frac{\alpha_2(s)^2}{2} E\bigg) \\ \nonumber & ~~~~~~~~~~\exp\{-i \alpha_2(s) P\} \exp\{-i \alpha_3(s) Q\}\exp\{-i \alpha_4(s)H\} \\ \nonumber & ~~
     +\alpha_4'(s) \alpha_3(s) P U(s) + \alpha_4'(s) \frac{\alpha_3(s)^2}{2} E U(s) \\ \nonumber
     &= \alpha_4'(s) H U(s) - \alpha_4'(s) \alpha_2(s) Q U(s) +\frac{1}{2}\alpha_4'(s) \alpha_2(s)^2 E U(s)  \\ \nonumber & ~~~~~~~+ \alpha_4'(s) \alpha_3(s) P U(s) + \frac{1}{2}\alpha_4'(s) \alpha_3(s)^2 E U(s) \\
     &= \bigg(\alpha_4'(s) H - \alpha_4'(s) \alpha_2(s) Q +\frac{1}{2}\alpha_4'(s) \alpha_2(s)^2 E \nonumber \\ & ~~~~~~~+ \alpha_4'(s) \alpha_3(s) P + \frac{1}{2}\alpha_4'(s) \alpha_3(s)^2 E\bigg)U(s).
\end{align}

The above form of $dU(s)/ds$ is exactly what was required. Now we one 
simply match the coefficients of the corresponding generators on both 
sides and arrive at the following differential equations of the 
$\alpha_i$'s. The terms in \ref{LHSterms} can be written as:
\begin{align}
\nonumber
    {\rm Term 1} &= \bigg(\alpha_4'(s) H  - \alpha_4'(s) \alpha_2(s) Q +\frac{1}{2}\alpha_4'(s) \alpha_2(s)^2 E  \\  & ~~~~~~~+ \alpha_4'(s) \alpha_3(s) P + \frac{1}{2}\alpha_4'(s) \alpha_3(s)^2 E\bigg) U(s), \\
    {\rm Term 2} &= \bigg(\alpha_3'(s) Q - \alpha_3'(s) \alpha_2(s) E\bigg)U(s), \\
    {\rm Term 3} &= \alpha_2'(s) P~ U(s), \\
    {\rm Term 4} &= \alpha_1'(s) E ~ U(s).
\end{align}

In consequence, Eq. \ref{LHSterms} can be written as: 
\begin{align}
\label{dudsfinal}
\nonumber
    \frac{dU(s)}{ds} &= -i \bigg[\alpha_4'(s) H  - \alpha_4'(s) \alpha_2(s) Q +\frac{1}{2}\alpha_4'(s) \alpha_2(s)^2 E  + \alpha_4'(s) \alpha_3(s) P \\ \nonumber & ~~~~~~~+ \frac{1}{2}\alpha_4'(s) \alpha_3(s)^2 E + \alpha_3'(s) Q - \alpha_3'(s) \alpha_2(s) E + \alpha_2'(s) P+ \alpha_1'(s) E  \bigg] U(s) \\ \nonumber
    & = -i \bigg[\alpha_4'(s) H+ \bigg\{\alpha_4'(s)\alpha_3(s) + \alpha_2'(s)\bigg\}P + \bigg\{-\alpha_4'(s)\alpha_2(s) + \alpha_3'(s)\bigg\}Q  \\
    & ~~~~~+\bigg\{ \alpha_1'(s) - \alpha_3'(s) \alpha_2(s) + \frac{1}{2} \alpha_4'(s) \alpha_3(s)^2 + \frac{1}{2} \alpha_4'(s) \alpha_2(s)^2\bigg\} E  \bigg] U(s).
\end{align}

Equating the coefficients of the generators from \ref{dudsfinal} and 
\ref{RHSapp}, we get the following: 
\begin{align}
    \alpha_1'(s) &= \alpha_2(s) v_P \sin (s v_Q)+\alpha_2(s) v_Q \cos (s v_Q)+\frac{\alpha_2(s)^2 v_Q}{2}-\frac{\alpha_3(s)^2 v_Q}{2},\\
    \alpha_2'(s) &= v_P \cos (s v_Q)-v_Q \sin (s v_Q)+\alpha_3(s) (-v_Q), \\
    \alpha_3'(s) &= v_P \sin (s v_Q)+v_Q \cos (s v_Q)+\alpha_2(s) v_Q,\\
    \alpha_4'(s) &= v_H.
\end{align}

The solutions of the above equations are as follows:
\begin{align}
\nonumber
    \alpha_1(s) &= \frac{1}{2} \bigg(\frac{1}{2} \sin (2 s v_Q) \left(C_1^2+2 s (C_1 v_P-C_2 v_Q)-C_2^2+s^2 \left(v_P^2-v_Q^2\right)\right) \\ & ~~~~~~~~+(C_1+s v_P) (C_2+s v_Q) \cos (2 s v_Q)+s (C_1 v_Q-C_2 v_P)\bigg)+C_3,\\
    \alpha_2(s) &= C_1 \cos (s v_Q)-C_2 \sin (s v_Q)+s v_P \cos (s v_Q)-s v_Q \sin (s v_Q),\\
    \alpha_3(s) &= C_1 \sin (s v_Q)+C_2 \cos (s v_Q)+s v_P \sin (s v_Q)+s v_Q \cos (s v_Q),\\
    \alpha_4(s) &= s v_H+C_4.
\end{align}

Imposing the boundary condition $U(s=0)= \mathbb{I}$, the constants $
C_i$'s can be fixed to be: 
\begin{align}
    C_1 =0, ~~~ C_2=0, ~~~ C_3 =0, ~~~ C_4=0.
\end{align}
which simplifies the solution as follows: 
\begin{align}
    \alpha_1(s) &= \frac{1}{4} s^2 \left(\left(v_P^2-v_Q^2\right) \sin (2 s v_Q)+2 v_P v_Q \cos (2 s v_Q)\right),\\
    \alpha_2(s) &= s (v_P \cos (s v_Q)-v_Q \sin (s v_Q)), \\
    \alpha_3(s) &= s (v_P \sin (s v_Q)+v_Q \cos (s v_Q)),\\
    \alpha_4(s) &= s v_H.
\end{align}

Now let us compute the complexity of the displacement operator 
as an illustration. The displacement operator in terms of 
the generators of the Harmonic oscillator group has been written 
in Eq. \ref{displacementoperator}: 
\begin{align}
    U_{\rm target}= \hat{D}= \exp(\sqrt{2}{\rm Re}(\alpha) Q+ \sqrt{2}{\rm Im}(\alpha) P).
\end{align}

Looking at $U_{\rm target}$, we immediately realize that it is 
not given in the form that we desire i.e., we have to disentangle 
the operator and express it as a product of the exponentials of 
the generators. Since the generators of Q and P commute with 
the commutator of Q and P (which is E), it is not difficult 
to see that $\hat{D}$ can be written as: 
\begin{align}
\nonumber
    \hat{D} &= \exp(\sqrt{2}{\rm Im}(\alpha) P+\sqrt{2}{\rm Re}(\alpha) Q ) \\ \nonumber & = \exp(\sqrt{2}{\rm Im}(\alpha)P)\exp(\sqrt{2}{\rm Re}(\alpha)Q)\exp(-{\rm Re}(\alpha){\rm Im}(\alpha)[P,Q]) \\
    &= \exp(\sqrt{2}{\rm Im}(\alpha)P)\exp(\sqrt{2}{\rm Re}(\alpha)Q)\exp(-{\rm Re}(\alpha){\rm Im}(\alpha)(-i E)).
\end{align}

In the above derivation, we have used the fact that if two 
operators $A$ and $B$ commutes with the commutator of $A$ 
and $B$, then we can write (from the BCH formula):
\begin{align}
    \exp(A+B)= \exp(A)\exp(B)\exp(-\frac{1}{2}[A,B]).
\end{align}

However, before comparing $\hat{D}$ with $U(s=1)$, we have to 
take into consideration the ordering of the generators, which 
should be identical to the ordering used while defining the 
generic element. In the generic element written in \ref{genericHOapp}, 
we see that the generator $E$ was at the beginning. Hence we have 
to reorder the exponentials in $\hat{D}$. Since $E$ commutes 
with both $Q$ and $P$, it can be shifted. Hence, $\hat{D}$ can 
be written as: 
\begin{align}
    \hat{D}= \exp(i{\rm Re}(\alpha){\rm Im}(\alpha) E)\exp(\sqrt{2}{\rm Im}(\alpha)P)\exp(\sqrt{2}{\rm Re}(\alpha)Q).
\end{align}

Therefore, we get the following:
\begin{align}
\nonumber
     U(s=1) &= \hat{D} \nonumber \\
     &= \exp(-i \alpha_1(1) E)\exp(-i \alpha_2(1) P)\exp(-i \alpha_3(1) Q)\exp(-i \alpha_4(1) H)  \\ &= \exp(i{\rm Re}(\alpha){\rm Im}(\alpha) E)\exp(\sqrt{2}{\rm Im}(\alpha)P)\exp(\sqrt{2}{\rm Re}(\alpha)Q),
\end{align}
from which we obtain the four conditions:
\begin{align}
    \alpha_1(1)= - {\rm Re}(\alpha){\rm Im}(\alpha), ~~~~ \alpha_2(1)= i \sqrt{2}{\rm Im}(\alpha), ~~~~ \alpha_3(1)= i \sqrt{2}{\rm Re}(\alpha), ~~~~ \alpha_4(1)= 0.
\end{align}

From the above conditions, we find that:
\begin{align}
    v_H=0, ~~~ v_P= i\sqrt{2}{\rm Im}(\alpha), ~~~ v_Q= i \sqrt{2}{\rm Re}(\alpha).
\end{align}

Therefore, the complexity of the displacement operator can be written a: 
\begin{align}
    C[\hat{D}]= \sqrt{v_P^2+v_Q^2+v_H^2}= 2|\alpha|,
\end{align}
which is exactly what we found in Sec. \ref{SectionDisplacementOperator}. 

\section{Commutation relations of the generators associated with the coupled harmonic oscillator}

As discussed in the main text (see Sec. \ref{sec:CoupledHO}), a basis 
of generators associated with a coupled harmonic oscillator that forms 
a closed Lie algebra are: 
\begin{align}
T_1&=H_1 = \frac{1}{2}(Q_1^2+P_1^2), \\
T_2&=H_2= \frac{1}{2}(Q_2^2+P_2^2), \\
T_3&=\frac{1}{2}(Q_1^2-P_1^2), \\
T_4&=\frac{1}{2}(Q_2^2-P_2^2), \\
T_5&=(Q_1P_1+P_1Q_1), \\
T_6&=(Q_2P_2+P_2Q_2), \\
T_7&=(Q_1Q_2+P_1P_2), \\
T_8&=(Q_1P_2+P_1Q_2), \\
T_{9}&=(Q_1Q_2-P_1P_2), \\
T_{10}&=(Q_1P_2-P_1Q_2). 
\end{align}

The commutation relations satisfied by the generators are as follows:
\begin{align*}
     & [T_1,T_2]=0,\quad [T_1,T_3]=-i T_5,\quad [T_1,T_4]=0, \quad [T_1,T_5]=i T_{3}, \quad [T_1,T_6]=0, \\
     & [T_1,T_8]=i T_{9}, \quad [T_1,T_{10}]= -i T_7,\quad [T_1,T_7]=i T_{10},\quad [T_1,T_{9}]= -i T_8,  \\
     & [T_2,T_3]=0, \quad [T_2,T_4]=-i T_{6}, \quad [T_2,T_7]=-i T_{10}, \quad [T_2,T_9]=-i T_{8}, \\
     & [T_2,T_8]=i T_{9},\quad [T_2,T_{10}]=i T_{7},\quad [T_2,T_6]=i T_{4},\quad [T_2,T_5]=0, \\
      & [T_3,T_{4}]= 0, \quad [T_3,T_5]=i T_{1}, \quad [T_4,T_5]=0, \quad[T_3,T_6]=0 , \quad [T_4,T_6]=i T_2, \\
      & [T_3,T_7]=i T_8, \quad [T_4,T_7]=i T_8, \quad [T_3,T_8]=i T_7, \quad [T_4,T_8]=iT_7, \\
      & [T_3,T_9]=-i T_{10}, \quad [T_4,T_9]=i T_{10}, \quad [T_3,T_{10}]=-i T_{9}, \quad [T_4,T_{10}]=iT_9, \\
      & [T_5,T_{1}]=-i T_{3}, \quad [T_6,T_{2}]=-iT_4, \quad [T_5,T_{3}]=-i T_{1}, \quad [T_6,T_{4}]=-iT_2, \quad [T_5,T_6]=0, \\
      & [T_5,T_{7}]=-2i T_{9}, \quad [T_6,T_{7}]=-2iT_9, \quad [T_5,T_{8}]=-2i T_{10}, \quad [T_6,T_{8}]=-2iT_{10}, \\
      & [T_5,T_{9}]=-2i T_{7}, \quad [T_6,T_{9}]=-2iT_{7}, \quad [T_5,T_{10}]=-2i T_{8}, \quad [T_6,T_{10}]=-2iT_{8}, 
\end{align*}

\begin{align*}
    & [T_7,T_1]= -i T_{10}, \quad [T_7,T_2]= i T_{10}, \quad [T_7,T_3]= -i T_8,\quad [T_7,T_4]=-i T_8, \\
    & [T_7,T_5]= 2 i T_9, \quad [T_7,T_6]= 2i T_9, \quad [T_7,T_8]= 2 i T_3+2i T_4, \quad [T_7,T_9]= -i T_6 -i T_5, \\
    & [T_7,T_{10}]= 2i T_1- 2i T_2, \quad [T_8,T_9]= -2 i T_1-2 i T_2, \quad [T_8,T_{10}]=-i T_6+i T_5, \\
    & [T_9,T_{10}]= 2 i T_3-2 i T_4.
\end{align*}

\end{appendix}

\bibliography{references}
\bibliographystyle{utphys}

\end{document}